\newcommand{\be}{\begin{equation}}\newcommand{\ee}{\end{equation}}
\newcommand{\bea}{\begin{eqnarray}}\newcommand{\eea}{\end{eqnarray}}
\newcommand{\nn}{\nonumber}\newcommand{\p}[1]{(\ref{#1})}
\newcommand{\lb}[1]{\label{#1}}

\def\sfrac#1#2{{\textstyle\frac{#1}{#2}}}

\documentclass[12pt]{article}
\usepackage{amsmath,amssymb}

\begin{document}
\begin{titlepage}
\begin{flushright}
\end{flushright}
\vspace{0.7cm}

\begin{center}
{\Large\bf Gauge Fields, Nonlinear Realizations,
 \vspace{0.2cm}

Supersymmetry}

\vspace{1.5cm}

{\large\bf
E.A. Ivanov} \\
\vspace{1cm}

 {\it Bogoliubov  Laboratory of Theoretical Physics, JINR,
141980 Dubna, Russia} \\
{\tt eivanov@theor.jinr.ru}

\end{center}
\vspace{1cm}

\begin{abstract}
\noindent This is a brief survey of the all-years research activity in the
Sector ``Supersymmetry'' (the former Markov Group)  at the Bogoliubov Laboratory of
Theoretical Physics. The focus is on the issues
related to gauge fields, spontaneously broken symmetries in the
nonlinear realizations approach, and diverse aspects of
supersymmetry.

\end{abstract}

\vspace{2.5cm}

\begin{center}
{\it To the memory of V.I. Ogievetsky and I.V. Polubarinov}
\end{center}
\end{titlepage}
\tableofcontents
\newpage

\section{Introduction}
The concepts which composed the Title of this paper lie in the ground of the modern mathematical theoretical
physics. From their very invention \cite{YM} - \cite{susy2}, they are constantly among the most priority directions of research in the Sector 3
of the Laboratory of Theoretical Physics. This Sector was originally named ``Markov Group'', after its first head, Academician
Moisei Alexandrovich Markov (1908 - 1994). Later on, for more than  20 years, it was headed by Professor
Victor Isaakovich Ogievetsky (1928 - 1996) and, since
the beginning of nineties, by the present author. The aim of the overview is to focus on the milestones of this long-lasting research
activity, with short explanations of their meaning  and significance for further worldwide developments of the relevant subjects. Besides
the studies concentrated around the Title issues, for the years passed since 1956 there were many considerable contributions
of the members of Sector 3 to other areas of theoretical physics, including the phenomenology of elementary particles,
the conceptual and mathematical basics of quantum mechanics, the renowned Ising model, etc. The choice of the topics of this overview
was determined by the preferences of the author and  the fact that his scientific interests always bore upon just these lines of investigations.

The structure of the paper is as follows. In Section 2 we deal with the period before invention of supersymmetry. Section 3 describes the most sound
results obtained in the domain of supersymmetry before the advent of Harmonic Superspace. The latter and related issues are
the subject of Section 4. In Section 5 we give a brief account of some other contributions of the Dubna group to the directions related to the Title.

Since many significant achievements go back to the pre-internet era, I describe them in some detail, with the hope that they could be
of interest for the modern generation of theorists.  This concerns the spin principle
(subsect. {\bf 2.1}), the ``notoph'' and ``inverse Higgs phenomenon'' (subsect. {\bf 2.2} and {\bf 2.6}), an interpretation of gravity and Yang-Mills
theories as nonlinear realizations (subsect. {\bf 2.5} and {\bf 2.7}), the complex superfield geometry of ${\cal N}=1$ supergravity (subsect. {\bf 3.3}),
the relation between the linear (superfield) and nonlinear (Volkov-Akulov) realizations of supersymmetry (subsect. {\bf 3.4}) and the whole
Section 4.

The present review partly overlaps with the review \cite{Dubna} which was devoted mainly to the supersymmetry
issues and so had a more narrow scope. Like in \cite{Dubna}, I apologize  for the inevitable incompleteness of the reference
list and a possible involuntary bias in my exposition of the investigations parallel to those performed in Dubna.

\section{Gauge fields, gravity and nonlinear realizations}
The first studies in the directions claimed in the title are dated by the beginning of sixties, and they were inspired by the invention
of non-abelian gauge fields by Yang and Mills in 1954 \cite{YM}. During a long time since its discovery, the Yang-Mills theory
was apprehended merely as a kind of elegant mathematical toy, since no any sign of non-abelian counterparts of the $U(1)$ gauge field,
 photon, was observed and nobody knew to which class of physical phenomena such a theory
could be applied. The situation has changed in the beginning of sixties after detection of strongly interacting massive vector bosons. It was
suggested that they can be analogs of photon for strong interactions and can be described by a mass-deformed Yang-Mills theory, with the minimally
broken gauge invariance\footnote{Now we know that the genuine theory of strong interactions is the quantum chromodynamics which
is the Yang-Mills theory for exact gauged ``color'' $SU(3)$ symmetry, with massless gluons as the relevant gauge fields. The second cornerstone of
the ``standard model'' is the electroweak theory which is the Yang-Mills theory for the gauge group $U(2) = SU(2)\times U(1)$,
with the photon and the triplet
of intermediate vector bosons as the gauge fields. The intermediate bosons are massive on account of the Brout-Englert-Higgs effect within a linear realization of
the spontaneous breaking of $U(2)$ symmetry. This mechanism of appearance of mass of the gauge fields does not break gauge invariance and preserves
the remarkable property of Yang-Mills theory to be renormalizable, like quantum electrodynamics.}.\\

\noindent{\bf 2.1 Spin principle.} The sharp growth of interest in non-abelian gauge
theories motivated Victor Isaakovich Ogievetsky
and Igor Vasil'evich Polubarinov (1929 - 1998) (under the approval and support of M.A. Markov) to carefully elaborate  on the nature
and role of gauge fields.
In the brilliant papers \cite{OP, OP1, LecOP} they put forward the so called ``spin principle'' as the basis of gauge theories.
Namely, they showed that
requiring one or another non-zero spin of field to be preserved in the interacting theory uniquely fixes the latter as a gauge
theory and the field with the
preserved spin as the relevant gauge field. The requirement of preservation of the spin 1 by the massless vector field
uniquely reproduces Maxwell theory in
the abelian $U(1)$ case and Yang-Mills theory in the case of few vector bosons \cite{OP,OP1}. Analogously, the theory
of self-interacting massless spin 2 field proved to be just
the Einstein gravity (treated as a field theory in Minkowski space-time) \cite{OP2}. It is the relevant gauge invariances
that ensure the neutralization
of superfluous spins which the gauge field can carry (spin 0 in the vector field, spins 0 and 1 in the tensor field, etc).
Moreover, the spin principle
applied to the theories with the gauge invariance broken by the mass terms fixes the latter in such a way that the ``would-be''
gauge field proves to be
coupled to a conserved current, and this condition ensures the preservation of the given spin in the massive case as well.

The spin of interacting fields was the pioneer concept introduced by Ogievetsky and Polubarinov. Before their papers,
it was a common belief that the quantum numbers of mass and spin characterizing the irreducible representations of
the Poincar\'e group are applicable only to the free particles and on shell (with the evident substitution of the notion of helicity for
that of spin for the massless particles).  Ogievetsky and Polubarinov were first to realize that the spin square  Casimir operator
of the Poincar\'e group $C_{(2)}$ can equally be defined for the interacting fields,  as opposed to the mass square
operator\footnote{Hereafter, our conventions are as in the book \cite{book}: $\eta^{mn} = {\rm diag} (1, -1,-1,-1)\,, \;\varepsilon_{0123} = 1\,, \; m, n = 0, 1, 2, 3\,$.}
$C_{(1)} = P^mP_m$  which cannot take any definite value on the interacting fields.

As an instructive example, we consider the case of vector field. The Casimir $C_{(2)}$ obtained as the square
of the Pauli-Lubanski vector (divided by $P^2\neq 0$
for further convenience) is in general expressed as
\be
C_{(2)} = \frac12 S^{mn}S_{mn} - \frac{1}{P^2}\,S^{mn}S^{q}_{\;\;n}P_m P_q\,,
\ee
where $S^{mn} = - S^{nm}$ is the matrix (spin) part of the full Lorentz generator $J^{mn} = S^{mn} + L^{mn}\,, \;
{\rm with} \; L^{mn} = i (x^m\partial^n - x^n\partial^m)\,,$ and $P_m = \frac{1}{i} \partial_m\,$. The operator $P^2 = - \Box$ does not take any definite value
in the theory with interaction,
$P^2 \neq 0$.  For the vector field $b_m^{i}\,,$ where $i$ is an index of some internal symmetry,  we have $(S^{pq})_{m}^{\,\,\,\,n} =
i(\delta^p_m \eta^{qn} - \delta^q_m \eta^{pn})$ and
\be
C_{(2)}\,b_m^i = C_{(2) m}^{\quad \;\, n}\,b^i_n = 2\Big[ b_m^i - \frac{1}{\Box}\partial_m (\partial^n b_n^i) \Big],
\ee
{\it i.e.} we obtain that the field $b_m^i$ is not an eigenfunction of  $C_2$. Let us decompose $b_m^i$ as
\be
b^i_m = {\bf b}^i_m + \partial_m \phi^i\,, \qquad  {\bf b}^i_m := (\delta_m^n - \frac{1}{\Box}\partial_m \partial^n)b^i_n\,, \quad
\phi^i := \frac{1}{\Box}(\partial^nb_n^i)\,. \lb{Dec}
\ee
It is easy to see that
\be
C_{(2)}\, {\bf b}^i_m = s(s+1)\,{\bf b}^i_m\,, \; s =1\,; \qquad C_{(2)}\, \partial_m \phi^i = 0\,.
\ee
Thus \p{Dec} is the decomposition of the vector field  $b_m^i$ into the transverse (spin 1) part ${\bf b}^i_m $ and the longitudinal (spin 0)
part $\partial_m \phi$. The question was how to arrange a theory in such a way that the vector field carries only spin 1 in the case
of non-trivial interaction.

Ogievetsky and Polubarinov started from a general Lagrangian for the massive vector fields interacting with themselves and some matter
fields $\Psi^A$, where $A = 1, 2, \ldots $ is an index of the same internal symmetry as for $b_m^i$,
\bea
&& L(b, \Psi) =  -\frac14 F^{mn\,i}F_{mn}^i + \frac12 m^2 b^{m i} b_m^i + L_{int}(b, \Psi) + L_{free}(\Psi)\,, \lb{LagrSpin1} \\
&& F_{mn}^i =
\partial_m b_n^i - \partial_n b_m^i\,. \nonumber
\eea
The equations of motion for $b_m^i$ read
\be
\partial^mF_{mn}^i + J^i_n + m^2 b_n^i = 0\,, \quad J^i_n := \frac{\partial L_{int}}{\partial b_n^i} -
\partial_p \frac{\partial L_{int}}{\partial (\partial_p b_n^i)}\,, \lb{EqSpin1}
\ee
where it is assumed that  $L_{int}(b, \Psi)$ does not include higher-order derivatives of $b_m^i$. As the necessary and sufficient
condition for $b_m^i$ to possess only spin 1 in the interacting theory, Ogievetsky and Polubarinov
rigorously proved that the equations of motion (including those for $\Psi^A$) should imply
\be
m^2\, \partial^n b_n^i = 0\,. \lb{CondSpin1}
\ee
This condition works  for both the massive and the massless cases. If $m^2 \neq 0$, one has $\partial^n b_n^i = 0$ which just means that
$C_{(2)}\,b_n^i = 2 b_n^i$, {\it i.e.} $b_n^i$ carries only spin 1. At $m=0$ \p{CondSpin1} is satisfied at any $\partial^m b_m^i$, which means
that the latter quantity is arbitrary and so is not physical. Its arbitrariness is ensured by the gauge invariance which
is thus the device to make $b_m^i$ to carry only spin 1 in the massless case. One can always choose the gauge $\partial^m b_m^i = 0$
which implies that $b_n^i  =  {\bf b}_n^i\,, \;  \partial^n{\bf b}_n^i = 0\,$, and so only spin 1 is really carried by the interacting
$b_n^i$ (this is true of course in any gauge).

The condition \p{CondSpin1} amounts to the conservation of the current  $J^i_n$ defined in \p{EqSpin1},
\be
\partial^n J^i_n = 0\,,
\ee
which means that the spin 1 fields $b_m^i$ couple to the conserved current. Using only this property, Ogievetsky and Polubarinov  were able
to uniquely  restore the interaction Lagrangian $L_{int}$ in \p{LagrSpin1}. Together with the free $F$ Lagrangian in \p{LagrSpin1} and modulo
the extra fields $\Psi^A$, this Lagrangian  is reduced in the general case to a sum of Yang-Mills Lagrangian for a semi-simple gauge group
and a number of the abelian $U(1)$ Lagrangians,  such that the dimension $d_{\{i\}}$ of the variety where the indices $i$ take their values
equals to the dimension
of the adjoint representation of the Yang-Mills gauge group plus the dimension of the abelian factors. For instance, if $d_{\{i\}} = 2$,
only $U(1)\times U(1)$ gauge group is possible, if  $d_{\{i\}} = 3$, the gauge group is either $SU(2)$ or $[U(1)]^3$, etc\footnote{It is
assumed that the gauge groups contain no solvable factors.}. For the fields $\Psi^A$ there naturally arise minimal gauge-invariant couplings to
the fields  $b_m^i$. These structures are also uniquely fixed from the requirement that  $b_m^i$ are coupled to the conserved current.

To be more precise, the solution obtained by Ogievetsky and Polubarinov (ignoring trivial abelian factors and the matter fields $\Psi$) is
\be
L (b) = -\frac1{4g^2} G^{mn\,i}G_{mn}^i + \frac12 m^2 b^{m i} b_m^i\,, \quad G_{mn}^i = \partial_m b_n^i - \partial_n b_m^i - c^{ilt}b_m^l b_n^t\,,
\ee
where $g$ is a coupling constant, $c^{ilt}$ are the structure constants of some semi-simple gauge group, with the hermitian generators $T^i$
satisfying the algebra $[T^i, T^l] = ic^{ilt}T^t\,$.
Ignoring the mass term, this Lagrangian is invariant under the gauge transformations with an arbitrary parameter $\lambda^i(x)$:
$\delta b^i_n = -\partial_n\lambda^i + c^{ikl}\,b_n^k\lambda^l\,$. So in the limit $m=0$ it becomes the standard
massless gauge-invariant Yang-Mills Lagrangian.
The mass term breaks the gauge invariance, but still retains the most important property of $b_n^i$ to be coupled to the conserved current.

It is interesting that the approach based on the requirement of preservation of spins 1 in the interaction does not assume in advance any gauge group,
the latter naturally arises, when revealing the structure of $L_{int}$ from this requirement, as the invariance group of
the full Lagrangian constructed in this way, modulo the mass term. The systematic use of the condition of coupling of $b_m^i$
to the conserved vector currents plays the crucial role in this derivation of the Yang-Mills Lagrangians (both massless and massive)
from the spin principle.

The same machinery was used in \cite{OP2} to derive the Einstein theory as a theory of symmetric tensor field carrying the spin 2
in interaction\footnote{To be more exact, in this case there is an admixture of spin 0. The pure spin 2 in interaction
corresponds to the conformal gravity.}.
Ogievetsky and Polubarinov showed that the Einstein-Hilbert Lagrangian can be consistently derived by requiring the spin 2 field
to be coupled to the conserved tensor
current. Once again, not only the massless Lagrangian with the exact {\it Diff}$\,R^4$ gauge symmetry can be restored in this way,
but also the appropriate
mass deformations thereof. In both cases, the crucial role was played by the requirement of coupling to the conserved current. It is worth noting that
the paper \cite{OP2} was one of the first papers where the gravity theory was treated on equal footing with other gauge theories as a field theory in
the flat background space-time. It is distinguished by the property of preservation of the definite spin 2 in interaction, quite analogously
to the treatment
of Yang-Mills theory as a field theory of definite spin 1 in interaction. Now such a treatment of gravity theories,
as well as supergravities, is of common
use.

The spin principle-inspired  view of gauge invariance as just a way to ensure a definite spin of the interacting field
proved to be very fruitful for further developments of gauge theories, including supergravity which is the unique self-consistent theory of interacting
gauge fields of the spin 2 (graviton) and spin 3/2 (gravitino). Actually, in the lectures \cite{LecOP} Ogievetsky and Polubarinov
have posed the question as to
what could be the gauge theory in which the Rarita-Schwinger field carries spin 3/2 in the interacting case. They made serious efforts
to find an answer \cite{Pcom}, but failed because nobody was aware of supersymmetry that time.

The careful analysis of how the spin principle is obeyed in the course of quantization, on the examples of quantum electrodynamics
and the theory of massive neutral gauge field, was accomplished by I.V. Polubarinov in the remarkable review \cite{IVPrev}\footnote{This
review was originally published in Russian as the preprint JINR-P-2421(1965).}. There,
also a comparative detailed  description of various approaches to quantizing the electrodynamics, including the historically  first ones,
was presented. \\

\noindent{\bf 2.2 Notoph.} While thinking on the group-theoretical grounds of gauge theories, Ogievetsky and Polubarinov discovered
a new gauge theory,
the gauge field of which is an
antisymmetric rank two tensor field still propagating spin 1 off shell and describing on shell a massless particle with zero helicity, the ``notoph'' \cite{notoph}.
Later on, the notoph was
re-discovered by Kalb and Ramond \cite{CaRa}. Now such gauge fields yielding an alternative off-shell description
of zero spin, as well as their higher-rank $p$-form generalizations,  are necessary ingredients of diverse
superstring and supergravity theories.

It is instructive to dwell on the notoph theory in some detail. It is described by the following Lagrangian
\be
L = -\frac12 A^m A_m + L_{int}(f_{mn}, \ldots)\,, \quad A^m := \frac12 \varepsilon^{mnpq}\partial_nf_{pq} \;
\Longleftrightarrow \; \partial_m A^m = 0\,. \lb{notophLag}
\ee
The antisymmetric tensor field $f_{mn}$ is the notoph gauge potential, it possesses the following gauge transformation law
\be
\delta f_{mn} = \partial_m\lambda_n - \partial_n\lambda_m\,, \lb{notophGT}
\ee
where $\lambda_m(x)$ is an arbitrary vector gauge parameter. The vector $A^m = \frac12 \varepsilon^{mnpq}\partial_nf_{pq}$ is the relevant
gauge invariant field strength and the condition $\partial_mA^m = 0$ is the corresponding Bianchi identity. The equation of motion for $f_{mn}$ reads
\bea
&& \frac12 \varepsilon^{mnst}\partial_s A_t = - J^{mn} \qquad {\rm or} \qquad
\Box f^{mn} - \partial^m\partial_p f^{pn} + \partial^n\partial_pf^{pm} = 2J^{mn}\,, \lb{EqNot} \\
&& J^{mn} := \frac{\partial L_{int}}{\partial f_{mn}}\,.
\eea
For the compatibility of the left-handed and right-handed parts of \p{EqNot} the tensor current $J_{mn}$ should be conserved,
\be
\partial^mJ_{mn} = 0\,. \lb{ConsNot}
\ee

To see how many on-shell degrees of freedom the gauge field $f_{mn}$ carries, one should take into account
that the gauge freedom \p{notophGT} actually involves
three independent gauge parameters because of the additional freedom  $\lambda _m \rightarrow \lambda_m + \partial_m\lambda$. So the field
$f_{mn}$ involves three independent off-shell degrees of freedom, like an abelian gauge field, and so represents spin 1 off shell.
On shell, two additional degrees of freedom are eliminated
by two analogs of the Gauss law in
electrodynamics
\be
\Delta f^{0b} + \partial_0 (\partial_af^{ab}) + \partial^b(\partial_af^{0a}) = -2J^{0b}\,,\quad
\Delta = -\partial_a\partial^a = \partial_a\partial_a\,,
\quad (a,b = 1,2,3)\,.
  \lb{Gauss}
\ee
To be convinced that this relation indeed amounts to the two independent equations, one can check that, in virtue
of the conservation law \p{ConsNot}, $\partial_aJ^{0a} = 0$,
only the transverse part $f_{tr}^{0b}$ of $f^{0b}$, $\partial_bf_{tr}^{0b} =0\,,$  gives contribution to \p{Gauss}.
As the result, we conclude that $f^{mn}$ indeed
comprises only one degree of freedom on shell.

An alternative way to demonstrate  that the notoph presents just another description of massless particle with zero helicity
is to perform the duality transformation relating the notoph theory to the theory of a single scalar field. For simplicity we limit ourselves
to the free theory, $L_{int} = 0$, and modify the free Lagrangian in \p{notophLag} by adding to it,
with the Lagrange multiplier $\varphi$, the 4-divergence
$\partial_mA^m$ going to become the Bianchi identity:
\be
L_0 = -\frac12 A^m A_m \; \Longrightarrow \; L_{dual} =  -\frac12 A^m A_m + \varphi \partial_mA^m\,. \lb{Dualiz}
\ee
When varying $L_{dual}$ with respect to $\varphi$, we obtain the Bianchi identity $\partial^mA_m=0$, after solving
which through $f^{mn}$ as in \p{notophLag}, the free
Lagrangian of notoph is recovered. On the other hand, eliminating $A^m$ from \p{Dualiz} by its algebraic
equation of motion, $A_m = -\partial_m\varphi$,
we obtain the free kinetic Lagrangian of the scalar field $\varphi$
\be
L_{dual}\; \Longrightarrow \;L_{\varphi} = \frac12 \partial^m\varphi \partial_m\varphi\,.
\ee
Note that in the case of non-trivial $L_{int}$ this duality holds in a local way only if $L_{int}$ depends
on $f^{mn}$ through the covariant gauge field
strength. So in general the descriptions of the massless spin zero particle through the scalar field and through the notoph
field result in physically
non-equivalent theories. It is just the description by the antisymmetric gauge fields that naturally appears
in superstring theory and some extended supergravities.

One more interesting property of the notoph theory is that its massive version is equivalent to the massive deformation of the abelian $U(1)$ theory
and so describes 3 independent degrees of freedom on shell. Once again, for simplicity we will consider the case without interaction
and modify the free action of the notoph as
\be
L_0 = -\frac12 A^m A_m \; \Longrightarrow \; L_{(m)} =  -\frac12 A^m A_m - \frac14 m^2 f^{mn}f_{mn}\,,
\quad A^m = \frac12 \varepsilon^{mnpq}\partial_nf_{pq}\,. \lb{massNot}
\ee
The equation of motion \p{EqNot} is modified as
\be
\Box f^{mn} - \partial^m\partial_p f^{pn} + \partial^n\partial_pf^{pm} + m^2f^{mn} = 0\,.\lb{MassEq}
\ee
Now the notoph gauge invariance is broken. Instead, eq.  \p{MassEq} implies the transversality condition
$\partial_mf^{mn} = 0$. It is easy to see that
this equation actually amounts to three independent conditions (because $\partial_m\partial_nf^{mn} = 0$
is satisfied identically), thus demonstrating that
the massive $f^{mn}$ indeed propagates three independent degrees of freedom on shell. We can dualize \p{massNot} as
\be
L_{(m)} =  -\frac12 A^m A_m - \frac14 m^2 f^{mn}f_{mn}\;  \Longrightarrow \; L_{(m)}^{dual} = \frac12 A^m A_m   +
\frac12 A^m \varepsilon_{mnpq}\partial^n f^{pq}- \frac14 m^2 f^{mn}f_{mn}\,, \lb{massNotdual}
\ee
where now $A_m$ is treated as an independent auxiliary field. Varying with respect to $A_m$, we come back to the theory \p{massNot}.
On the other hand, varying with respect to $f^{mn}$, we obtain
\be
f_{mn} = -\frac{1}{m^2}\varepsilon_{mnpq}\partial^pA^q\,.
\ee
Substituting it into \p{massNotdual}, we obtain, up to a rescaling
\be
L_{(m)}^{dual}  = -\frac14 F^{mn}F_{mn} + \frac12 m^2 A^mA_m\,, \quad F_{mn} = \partial_m A_n - \partial_n A_m\,.
\ee

Thus both the theory of gauge abelian vector field describing on shell 2 degrees of freedom (helicities $\pm 1$)
and the gauge theory of notoph describing on shell
one degree of freedom (zero helicity), after the minimal mass deformation yield the same theory of massive spin 1
which propagates 3 degrees of freedom on shell.
So these two gauge theories are complementary to each other in the sense that the full set of helicities
of the relevant particles equals to the set of the projections of
the massive spin 1. So they can be treated as two different massless limits of the abelian massive spin 1 theory
\footnote{This complementarity does not generalize  to the non-abelian case, at least in a direct way.}.

Though notoph is a necessary mathematical ingredient of supergravities and string theory, it is still an open question whether
it could manifest itself  in a more phenomenological context as an elementary particle, like the standard vector gauge fields (photon, intermediate vector bosons, etc).
The authors of \cite{notoph} indicated a few possible processes where the notoph could be produced, but no any sign of it was detected
so far. Despite the fact that the standard model and its currently discussed generalizations have seemingly no direct need in such an entity,
nevertheless, to my knowledge, no any no-go theorem against such a possibility was adduced.\\

\noindent{\bf 2.3 Spinors in the gravitation theory.} One more important and far-reaching result
of the Ogievetsky-Polubarinov collaboration concerned the description of spinors
in general relativity. They showed \cite{spinors} that there is no direct necessity to introduce
the orthogonal repere (vierbein) in order to
construct the invariant coupling of Dirac fields to the gravitons; this can be done in a minimal way  by ascribing, to spinorial fields,
the nonlinear in graviton transformation law under the space-time diffeomorphism group, without introducing any extra entities. Actually,
the paper \cite{spinors} anticipated the nonlinear realization method which was discovered and applied for description of spontaneously broken symmetries in
the low-energy strong interactions (``chiral dynamics'') by Schwinger, Weinberg, Volkov and others in a few years. Also, it was the first step towards
interpreting gravity as a theory of two spontaneously broken space-time symmetries, the affine and conformal ones,
by Borisov and Ogievetsky \cite{BorOg}.\\

\noindent{\bf 2.4 Nonlinear realizations and chiral dynamics.} V.I. Ogievetsky, together with
his PhD student Boris Zupnik (1945 - 2015), took active
participation in developing and applying the above mentioned nonlinear realization
and effective Lagrangian methods of describing various low-energy systems. In particular, they proposed a new general method of
constructing nonlinear realizations of the groups $U(N)$ \cite{OgZu}, before the appearance of the seminal papers
on the general theory of nonlinear
realizations \cite{nonl12,nonl3}. Also, a new effective Lagrangian was proposed to describe the $(\pi, \rho, A_1)$ system,
with the maximally smooth momentum behavior
of the corresponding amplitudes \cite{OgZu2}. This model found a few interesting and unexpected applications,
in particular it was used to calculate contributions
of the so called exchange currents in some nuclear reactions \cite{Itu}.\\

\noindent{\bf 2.5 Einstein gravity from nonlinear realizations.} As a natural continuation of this research activity,
Ogievetsky was soon got interested
in applying the nonlinear realizations method, developed in \cite{nonl12}
basically for internal symmetries, to the space-time symmetry groups including the Poincar\'e group as a subgroup \cite{nonl3,OgCar}.
Studying the structure of the diffeomorphism group
in $R^4\,$, he discovered that this infinite-dimensional group can be nicely represented as a closure
of its two finite-dimensional subgroups, affine and conformal ones,
intersecting over the common Weyl subgroup (the semi-direct product of the Poincar\'e group and dilatations)\cite{Teorem}.
This remarkable observation is now known as
the Ogievetsky theorem. It has many applications, in particular, in supergravity and $M$-theory
(see, e.g., \cite{West} and refs. therein).

To explain the meaning of this theorem, let us write the special conformal transformation in the Minkowski space,
\be
\delta_\beta x^m =  \beta^m x^2 -2(\beta\cdot x) x^m\,, \lb{conf}
\ee
where $\beta^m$ is the transformation parameter of the mass dimension, and the general linear $gl(4,R)$ transformation
\be
\delta_\lambda x^m = -\Lambda^m_{\;\;n} x^n \,, \lb{aff}
\ee
where the constant dimensionless parameters $\Lambda^m_{\;\;n}$ form a real $4\times 4$ matrix comprising 16 independent parameters.
The transformations \p{aff} involve 6-parameter transformations with the antisymmetric matrix $\Lambda_{[mn]}$, which form
the Lorentz subalgebra $so(1,3)$ in $gl(4,R)$, and the 10-parameter transformations with the symmetric matrix $\Lambda_{(mn)}$,
which belong to the symmetric coset of the group ${GL}(4,R)$ over the Lorentz group ${SO}(1,3)$. One can also add the translations
of $x^m$ which complete
$gl(4,R)$ to the affine algebra $A(4)$. The translations, together with the Lorentz transformations
and dilatations, also extend \p{conf} to the 4-dimensional conformal algebra $so(4,2)$.

Let us now define the generators corresponding to \p{conf} and symmetric part of \p{aff}
\be
K_n := -{i}\left(x^2 \delta^m_n - 2 x^m x_n\right)\partial_m\,, \quad R_{st} := -i(x_s\partial_t + x_t \partial_s)\,,
\ee
and compute their commutator, $[K_n, R_{st}]$. In this commutator, besides the generator $K_n$, we find new generators
of the second order in $x^m$
\be
\sim \eta_{ns} x_t (x^p\partial_p)\,, \quad \sim x_s x_t \partial_n\,.
\ee
Commuting these new generators with $K_n$, $R_{st}$ and themselves, we encounter new generators of the third order in $x^m$, etc.
Ogievetsky has proved that this process does not terminate at any finite step and produces the whole set of generators
\be
L^{n_1 n_2 n_3 n_4}_m =  (x^0)^{n_1}(x^1)^{n_2}(x^2)^{n_3}(x^3)^{n_4}\partial_m\,,
\ee
constituting an infinite-dimensional diffeomorphism group\footnote{To be more exact, the connected subgroup
of  {\it Diff} $R^4$ consisting
of all transformations expandable in the Taylor series around the origin $x^m=0\,$.} {\it Diff} $R^4$
\be
\delta x^m = f^m(x) = \sum_{n_1,n_2, n_3, n_4} c^{\{n_1n_2n_3n_4\}\,m}(x^0)^{n_1}(x^1)^{n_2}(x^2)^{n_3}(x^3)^{n_4}\,.
\ee
Here $n_1, \ldots n_4$ are arbitrary non-negative integers, $n_i \geq 0\,$, and $c^{\{n_1n_2n_3n_4\}\,m}$ are constant parameters.

Based on this theorem, Ogievetsky with his PhD student Alexander Borisov constructed the sigma-model-type theory
invariant under the simultaneous nonlinear realizations
of the affine and conformal groups and showed that
this theory is nothing else as the Einstein gravitation theory \cite{BorOg}. Let us recall the basic details of their construction.

The starting point is two algebras, affine and conformal, involving, respectively, the generators $(P_m, L_{mn}, R_{mn})$ and $(P_m, K_n, L_{mn}, D)$
with the following commutation relations\footnote{We use slightly different conventions as compared to \cite{BorOg}.  They are the same
as in \cite{book}.}
\bea
&& [L_{mn}, L_{pq}] = i\Big(\eta_{np}L_{mq} - \eta_{mp}L_{nq} - (p \leftrightarrow q)\Big), \nn \\
&& [L_{mn}, R_{pq}] = i\Big(\eta_{np}R_{mq} - \eta_{mp}R_{n q} +
(p \leftrightarrow q) \Big), \nn \\
&& [R_{mn}, R_{pq}] = i\Big(\eta_{mp}L_{nq} +  \eta_{np}L_{mq} + (p \leftrightarrow q)\Big), \nn \\
&& [L_{mn}, P_q] = i\Big(\eta_{nq}P_m - \eta_{mq}P_n\Big), \quad
[R_{mn}, P_q] = i\Big(\eta_{nq}P_m + \eta_{mq}P_n\Big), \lb{affineAl} \\
&& [P_m, P_n] = [K_m, K_n] = 0\,, \quad [D, P_m] = iP_m, \quad  [D, K_m] = -iK_m\,, \nn \\
&& [L_{mn}, K_q] = i\Big(\eta_{nq}K_m - \eta_{mq}K_n\Big), \quad [P_m, K_n] = 2i\Big( \eta_{mn} D + L_{mn}\Big). \lb{confAl}
\eea
These two algebras intersect over the Weyl algebra $( P_s, L_{mn}, D = \frac12 R^m_{\;\;\;m})$.

As the next step, the authors of \cite{BorOg} constructed a nonlinear realization of the affine group ${\cal A}(4)$ in the coset space over
the Lorentz subgroup,
\be
\frac{{\cal A}(4)}{SO(1,3)} \quad \sim \quad \frac{\{P_m, L_{mn}, R_{mn}\}}{\{L_{mn}\}}\,,
\ee
with the following parametrization of the coset element
\be
G(x, h_{mn}) = {\rm e}^{ix^mP_m}{\rm e}^{\frac{i}{2} h^{mn}(x)R_{mn}}\,,\lb{affCos}
\ee
where $h^{mn}(x)$ is the symmetric tensor Goldstone field. Under the left multiplications by an element $g$ of ${\cal A}(4)$,
the coset representative is transformed as
\be
G'(x',h') = {\rm e}^{i{x}^m{}'P_m}{\rm e}^{\frac{i}{2} {h}^{mn}{}'(x')R_{mn}} = g\, G (x,h)\,{\rm e}^{-\frac{i}{2} u^{mn}(x,h,g)L_{mn}}\,, \lb{Left}
\ee
where $u^{mn}(x,h,g)$ is the induced Lorentz group parameter. In particular, left multiplications by ${\rm e}^{\frac{i}{2}\lambda^{[mn]}L_{mn}}$
and ${\rm e}^{\frac{i}{2}\lambda^{(mn)}R_{mn}}$ yield for $x^m$ the transformations \p{aff}, with $\Lambda^{mn} = \lambda^{[mn]} + \lambda^{(mn)}\,$.
According to the general prescriptions of nonlinear realizations, one can now construct the left-covariant Cartan one-forms
\be
G^{-1}dG = i\omega_{(P)}^m P_m + \frac{i}{2} \omega_{(R)}^{mn} R_{mn} + \frac{i}{2} \omega_{(L)}^{mn} L_{mn}\,.\lb{CartanF}
\ee

The form $\omega_{(P)}^m$ is calculated  to be
\bea
&& \omega_{(P)}^m = e^m_p dx^p\,, \quad e^m_p = ({\rm e}^{\frac{i}{2} h^{st} \hat{R}_{st}})^m_p = ({\rm e}^{h})^m_p
= \delta^m_p + h^m_p + \frac12 h_p^nh_n^m + \ldots\,,
\lb{Tetr} \\
&& (\hat{R}_{st})^m_p = -i(\eta_{sp}\delta^m_t + \eta_{tp}\delta^m_s)\,. \nonumber
\eea
The external product of four forms $\omega_{(P)}^m$ defines the invariant $R^4$ volume element, {\it Vol}$\,R^4
= \det e^m_p\,d^4x = {\rm e}^{h^m_m}\, d^4x$,
the form  $\omega_{(R)}^{mn}$
defines the covariant derivative of the tensor Goldstone field $h^{mn}$, $\omega_{(R)}^{mn} = \omega_{(P)}^s\nabla_s h^{mn}\,,$ and
the inhomogeneously transforming form  $\omega_{(L)}^{mn}$ - the covariant differential and the covariant derivative of the ``matter'' field $\Psi^A$
transforming by some irreducible representation of the Lorentz group with the matrix generators $(S_{mn})^A_{\;\;B}$
\bea
&&{\cal D}\Psi^A = d\Psi^A + \frac{i}{2}\omega_{(L)}^{mn} (S_{mn})^A_{\;\;B}\Psi^B := \omega_{(P)}^s{\cal D}_s\Psi^A\,,   \lb{CovDif} \\
&&{\cal D}_s\Psi^A = (e^{-1})_s^p\partial_p\Psi^A  + \frac{i}{2}{\cal V}^{mn}_s (S_{mn})^A_{\;\;B}\Psi^B\,, \quad
\omega_{(L)}^{mn} := \omega_{(P)}^s{\cal V}^{mn}_s\,. \lb{CovDer}
\eea
All these objects were explicitly computed in \cite{BorOg}. An important observation was that the
Lorentz connection in \p{CovDer}
can be generalized, without affecting its transformation properties, by adding three independent combinations
of the covariant derivatives $\nabla_ph^{mn}$
\bea
{\cal V}_{mn, s} \quad \Rightarrow \quad {\cal V}^{gen}_{mn, s} = {\cal V}_{mn, s} + \alpha_1\,\nabla_{[m} h_{n]s}
+ \alpha_2\,\eta_{s[m}\nabla_{n]} h^p_{\;\;p}
+ \alpha_3\, \eta_{s[m} \nabla_p h^p_{\;\;n]} \,.\lb{GenCon}
\eea

The next step was the analogous construction of nonlinear realizations of the conformal group in the coset with the same stability subgroup
\be
\frac{SO(2,4)}{SO(1,3)} \; \sim \; \frac{\{P_m, L_{mn}, K_n, D\}}{\{L_{mn}\}}\,, \qquad \tilde{G}(x,\sigma, \varphi)
= {\rm e}^{ix^mP_m}{\rm e}^{i\varphi^m(x)K_m}{\rm e}^{{i}\sigma(x)D}\,.\lb{confCos}
\ee
The conformal group is realized by left shifts on the coset element $\tilde{G}(x,\sigma, A)$. In particular, the left multiplication
by ${\rm e}^{i\beta^mK_m}$
generates for $x^m$ just the transformation \p{conf}. The left-covariant Cartan forms are defined by
\be
\tilde{G}^{-1}d\tilde{G} = i\tilde{\omega}_{(P)}^mP_m + i\omega_{(D)} D + i \omega^m_{(K)}K_m + \frac{i}{2}\tilde{\omega}_{(L)}^{mn}L_{n}\,.
\ee
They can be easily computed. In particular,
\be
\tilde{\omega}_{(P)}^m = {\rm e}^\sigma dx^m\,, \quad \omega_{(D)} = d\sigma - 2 \varphi_m dx^m\,, \quad \tilde{\omega}_{(L)}^{mn} = 2(\varphi^mdx^n - \varphi^ndx^m)\,.
\ee
Taking into account that $\sigma = \frac14 h^m_{\;\;m}$ (because of the identification $D = \frac12 R^m_{\;\;m}$),
we observe that the invariant volume {\it Vol}$\,R^4$ is
the same in both nonlinear realizations
$$
{Vol}\,R^4 ={\rm e}^{4\sigma} d^4 x = {\rm e}^{h^m_{\;\;m}} d^4 x\,.
$$
The vector Goldstone field $\varphi_m(x)$ is unessential, as it can be covariantly eliminated by equating to zero the Cartan form $\omega_{(D)}$
\be
\omega_{(D)} = 0\, \quad \Rightarrow \quad \varphi_m = \frac12 \partial_m \sigma
\ee
(this is a particular case of the ``inverse Higgs phenomenon'', see the next subsection). The covariant derivative of the ``matter''
fields is given by the expression
\be
\tilde{\cal D}_m\Psi^A = {\rm e}^{-\sigma}\Big( \partial_m \Psi^A - i\partial_n\sigma (S^{n}_{\;\;m})^A_{\;\;B}\Psi^B \Big).\lb{ConfCovDer}
\ee

As the last step in deriving the Einstein gravity, the authors of \cite{BorOg} analyzed the issue of simultaneous covariance
under both nonlinear realizations constructed.
As a consequence of the Ogievetsky theorem, the theory exhibiting such a covariance should be invariant under the full {\it Diff}$\,R^4$ group.

The ${\cal A}(4)$ Goldstone field $h^{mn}$ can be divided as $ h^{mn} = \hat{h}^{mn} + \frac14 \eta^{mn}{h^p_{\;\;p}} = \hat{h}^{mn} + \eta^{mn}\sigma$,
where the traceless tensor field $\hat{h}^{mn}$ can be treated as a ``matter'' field with respect to
the nonlinear realization of the conformal group. Then one requires that the ${\cal A}(4)$ covariant derivative \p{CovDer} with the generalized
Lorentz connection \p{GenCon} involves the dilaton field $\sigma$ and its derivatives only through the conformal covariant derivative \p{ConfCovDer}.
This requirement uniquely fixes the coefficients in \p{GenCon} as $\alpha_1 = -2, \alpha_2 = \alpha_3 =0\,$. The resulting covariant
derivative is simultaneously
covariant under the nonlinear realizations of both affine and conformal groups and, hence, under their closure {\it Diff}$\,R^4$.
Borisov and Ogievetsky also showed
that no combinations of the ${\cal A}(4)$ covariant derivatives $\nabla_sh^{mn}$ exist, such that they are covariant under the conformal group.
The latter
property corresponds to the well-known fact that in the Riemannian geometry no tensors involving the first derivative of the metric tensor
can be constructed.
The first non-trivial tensor contains two derivatives and in the formulation of \cite{BorOg} it is constructed as the covariant curl
of the Lorentz connection
${\cal V}_{mn, s}^{gen}$:
\be
({\cal D}_m{\cal D}_n - {\cal D}_m{\cal D}_n)\Psi^A = \frac{i}{2}{\rm R}_{mn}^{pq} (S_{pq})^A_{\;\;B}\Psi^B\,.
\ee
Since ${\rm R}_{mn}^{pq}$ undergoes induced Lorentz rotation with respect to all of its indices, the object ${\rm R} := {\rm R}_{mn}^{mn}$
is invariant
under the simultaneous nonlinear realizations of the affine and conformal groups and hence under the group {\it Diff}$\,R^4$.
It can be represented as the standard
Riemann scalar curvature with the metric
\be
g_{mn} = e_m^p e_{n p} = ({\rm e}^{ih^{sq}\hat{R}_{sq}})_m^{\;p}\,\eta_{np} = \eta_{mn} + 2 h_{mn} + \ldots \,,
\ee
having the standard transformation properties under the coordinate transformations, $\delta g_{mn} = -\partial_m\delta x^p g_{pn}
- \partial_n\delta x^p g_{mp}\,$.
The minimal invariant action coincides with the Einstein-Hilbert action
\be
-\frac{1}{16 \pi G}\int d^4 x \sqrt{ -g}\, {\rm R}\,,
\ee
where $G = \frac1{4\pi} f^2$ and the constant $f$, $[f] = 1$, arises as the result of the standard rescaling of the dimensionless
Goldstone field $h_{mn}$,
$h_{mn} \rightarrow f h_{mn}$. The couplings to matter fields are constructed using the covariant derivative \p{CovDer}
with the connection ${\cal V}_{mn, s}^{gen}$ (with the fixed coefficients ensuring the conformal group covariance).
The connection ${\cal V}_{mn, s}^{gen}$ can be related to the Christoffel symbols. The nonlinear transformation
law for spinors deduced in \cite{spinors}
immediately follows from the general transformation law of matter fields in the realization \p{Left}, with the induced Lorentz parameter,
\be
\delta\Psi^A = \frac{i}{2} u^{mn}(h)(S_{mn})^A_{\;\;B}\Psi^B\,,
\ee
upon specializing, e.g.,  to the $(1/2,0)$ spinor representation, with $\frac12 (\sigma_{mn})^\alpha_\beta$ as the spin part of the Lorentz generators.
In fact, the theory obtained in \cite{BorOg} can be reproduced from  the Einstein theory formulated in terms of vierbeins $e_m^a\,$
by gauge-fixing the local Lorentz rotations in the tangent space in such a way that the antisymmetric part of $e_m^a$ vanishes.

The work \cite{BorOg} turned out very important in the conceptual sense, because it exposed, for the first time,
the double nature of the spin 2 graviton field. On the one hand, it is a gauge field for the diffeomorphism group (modulo some subtle
issues, see subsect. {\bf 2.8}) and,
on the other, as follows from \cite{BorOg},
it is the tensorial Goldstone field accompanying the spontaneous breakdown of the finite-dimensional affine and conformal groups. Later on,
it was shown that any gauge field
can be interpreted as a Goldstone field associated with some infinite-dimensional global symmetry \cite{IvOg1} (see subsect. {\bf 2.7}).
The space-time diffeomorphisms are distinguished
in that they can be represented as a closure of two finite-dimensional groups, while there is no analogous theorem for the Yang-Mills
type gauge groups. As was shown in \cite{Iv}
and \cite{INied3}, the super Yang-Mills and supergravity theories (at least, the simple ${\cal N}=1, 4D$ ones) can also be reproduced from
the nonlinear realizations of some supergroups. The interpretation
of the Yang-Mills and graviton fields,
as well as their super Yang-Mills and supergravity counterparts,
as the Goldstone (super)fields, and the associate (super)gauge and (super)gravity theories as nonlinear realizations, posed a natural question
as to what could be {\it linear} realizations of these groups and which generalizations of {\it linear} sigma models of the underlying
spontaneously broken  symmetries
could  correspond to such realizations. Until now, there is no answer to this question. Since, in such hypothetical theories,
the Yang-Mills and/or gravity fields
should appear inside some linear multiplets, while the symmetry generators carry Lorentz indices, these multiplets should be infinite-dimensional and
include fields with higher spins. So these hypothetical theories should be a sort of higher-spin or string-like theories (M-theory?).
Note that in the nonlinear realization
formulation the space-time coordinate $x^m$ itself appears as a coset parameter. In the conjectured linear multiplets it should be present
on equal footing with those components which are going to become Goldstone fields after spontaneous breaking. \\

\noindent{\bf 2.6 Inverse Higgs phenomenon.} The classical Nambu-Goldstone theorem claims that, to any generator of spontaneously broken
symmetry in the quantum field theory, there should correspond a massless Goldstone field with the inhomogeneous transformation law
under this generator, such that it starts with the relevant transformation parameter.  The basic result of the paper \cite{InvH} was the observation
that in nonlinear realizations of space-time symmetries certain Goldstone fields can be covariantly traded for space-time derivatives
of some minimal set of such fields,
and there were established the general criterions under which this becomes possible. The condition under which the given Goldstone field
admits an  elimination is that the commutator of the space-time translation
generator with the corresponding spontaneously broken generator again yields a spontaneously broken generator. This phenomenon
was called ``Inverse Higgs phenomenon'' or ``Inverse Higgs effect''.
It proved to work with an equal efficiency in the superfield theories as well. Now it is of indispensable use in theories
with nonlinear realizations of space-time (super)symmetries.

As an example of inverse Higgs effect, let us reproduce the massive particle ($0$-brane) in
the flat $2D$ space-time by the nonlinear realization method applied to the $2D$ Poincar\'e group ${\cal P}_{(2)}$ \cite{Diverse}.
The latter involves two translation generators $P_0, P_1$ and the
$SO(1,1)$ Lorentz generator $L$, with the only two non-vanishing commutators
\be
[\,L, P_0\,] = iP_1~, \quad  [\,L, P_1\,] = i P_0~.\lb{LPPcomm}
\ee
Then we construct a nonlinear realization of ${\cal P}_{(2)}$, with the
one-dimensional ``Poincar\'e'' generator $P_0$ as the only one to which a coordinate (time) is associated as the coset parameter.
Two other generators pick up as the relevant parameters the ``Goldstone'' fields $X(t)$ and $\Lambda(t)$, giving rise to the following coset
element
\be
G = \mbox{e}^{itP_0} \, \mbox{e}^{iX(t)P_1}\, \mbox{e}^{i\Lambda(t)L}~.
\label{toycos}
\ee
The group ${\cal P}_{(2)}$ acts as left shifts of $G\,$, $G\rightarrow G' = \mbox{e}^{iaP_0}\mbox{e}^{ia_1P_1}\mbox{e}^{i\sigma L}G$.
The Cartan forms
\be
G^{-1}\,d G = i\,\omega_0\, P_0 + i\,\omega_1\,P_1 + i\omega_L\,L~,
\label{toyf}
\ee
\bea
&& \omega_0 = \sqrt{1 + \Sigma^2}\,dt + \Sigma \,dX~, \quad
\omega_1 = \sqrt{1 + \Sigma^2}\,dX + \Sigma \,dt~, \nn \\
&& \omega_L= {1\over \sqrt{1 + \Sigma^2}}\,d\Sigma~, \qquad
\Sigma \equiv \mbox{sh}\,\Lambda\,,
\label{om}
\eea
by construction are invariant under this left action. We observe that the Lorentz Goldstone field $\Sigma (t)$
can be traded for
$\dot{X}(t)$ by the inverse Higgs constraint
\be
\omega_1 = 0 \quad \Rightarrow \quad \Sigma =
- {\dot{X}\over \sqrt{1-\dot{X}^2}}~.
\label{ih}
\ee
This constraint is covariant since $\omega_1$ is the group invariant (in the
generic case, the coset Cartan forms undergo homogeneous rotation
in their stability subgroup indices). Thus the obtained expression
for $\Sigma$ possesses correct
transformation properties \footnote{The possibility to eliminate the field $\Sigma$ (or $\Lambda$) follows from the criterion of the elimination
mentioned earlier. Indeed, the commutator of the time-translation operator $P_0$ with the broken generator $L$ yields the broken
generator $P_1$ (see \p{LPPcomm}).}.
Substituting it into the remaining
Cartan forms we find
\be
\omega_0 = \sqrt{1 - \dot{X}^2}\,dt~, \quad \omega_L =
\sqrt{1 - \dot{X}^2}\,
\frac{d}{dt}\left(\frac{\dot{X}}{\sqrt{1 - \dot{X}^2}}\right)dt~.
\ee
The simplest invariant action, the covariant length
\be
S = \int \omega_0 = \int dt \sqrt{1 - \dot{X}^2}~,
\label{toyact}
\ee
is recognized, up to a renormalization factor of the dimension of mass,
as the action of $2D$ massive particle in the static gauge $X^0 (t) = t\,$.

Another text-book example of how the inverse Higgs phenomenon works is the derivation of the Alfaro-Fubini-Furlan conformal mechanics
from the non-linear realization of the $d=1$ conformal group $SO(2,1) \sim SU(1,1)$ \cite{ConDyn}.
Its important application in constructing non-linear realizations of $4D$ conformal group was already discussed in the previous subsection.
Its use is crucial for deducing the superfield actions of branes in the approach based on the concept of partial spontaneous
breaking of global supersymmetry (PBGS) (see, e.g., \cite{IHrecent} and references therein). In the PBGS models this effect has also
dynamical manifestations, giving rise, in some cases, to the equations of motion as the result of equating to zero some appropriate Cartan
forms (see, e.g. \cite{BIK}). Such an extended inverse Higgs effect was also applied for deducing some two-dimensional integrable
systems from  nonlinear realizations of (super)groups (see, e.g., \cite{N2L,IK1}) and deriving new kinds
of superconformal mechanics in the superfield approach \cite{IKL3}. Recently, it was applied for construction of the Galilean conformal
mechanics in \cite{Galil}.

The inverse Higgs phenomenon plays the key role in proving that any gauge theory, like the gravitation theory, admits an alternative interpretation
as a theory of spontaneous breakdown, with the gauge fields as the corresponding unremovable Goldstone fields \cite{IvOg1}.\\

\noindent{\bf 2.7 Yang-Mills theory as a nonlinear realization.} The basic idea of \cite{IvOg1} was to represent the Yang-Mills
gauge group\footnote{To be more exact, the connected component of the full gauge group, spanned by the gauge functions
admitting a decomposition into the Taylor series in a vicinity of $x^m =0\,$.} as a group with constant parameters
and an infinite number of tensorial generators.

One starts with some internal symmetry group with the generators $T^i$,
\be
[T^i, T^k] = i c^{ikl}T^l\,,\lb{Global}
\ee
and decomposes $\lambda^i(x)T^i$ as
\be
\lambda^i(x)T^i = \lambda^iT^i + \sum_{n\geq 1}\frac1{n!}\,\lambda^i_{m_1\cdots m_n} x^{m_1}\ldots x^{m_n}T^i\,.
\ee
Denoting $T^{(m_1 \cdots m_n)i} := x^{m_1}\ldots x^{m_n}T^i$, we indeed can rewrite the gauge parameter with values in the Lie algebra \p{Global} as
\be
\lambda^i(x)T^i = \lambda^iT^i + \sum_{n\geq 1}\frac1{n!}\,\lambda^i_{m_1\cdots m_n}T^{(m_1 \cdots m_n)i}\,,
\ee
{\it i.e.}  as a particular  representation of the infinite-dimensional algebra generated
by $T^{(m_1 \cdots m_n)i}$. Viewed as the abstract algebra, this set of  generators, together with the 4-translation generator
$P_m = \frac1{i}\partial_m\,$,  is closed under the commutation relations
\bea
&& [T^{(m_1 \cdots m_n)i}, T^{(p_1 \cdots p_s)k} = i c^{ikl}T^{(m_1\cdots m_n p_1 \cdots p_s) l}\,, \quad n\geq 1\,, \; s\geq 1\,, \nn \\
&& [T^i, T^{(m_1 \cdots m_n)k}] = i c^{ikl}T^{(m_1 \cdots m_n)l}\,, \nn \\
&& [P_n, T^{m\,i}] = -i \delta^m_n T^i\,, \lb{PTm} \\
&& [P_n, T^{(m_1 \cdots m_k)i}] = -i\Big( \delta^{m_1}_nT^{(m_2 \cdots m_k)i} + \ldots
+ \delta^{m_k}_n T^{(m_1 \cdots m_{k-1})i}\Big), \quad k \geq 2\,, \lb{InfDim}
\eea
to which one should add \p{Global}. With respect to the Lorentz group, the generators $T^{(m_1 \cdots m_n)i}, n\geq 1\,,$ form symmetric
tensors of the rank $n$. In fact, the Lorentz group can be considered as external automorphisms of the algebra of the remaining
generators and so it decouples.

The full infinite-dimensional group involving the abstract generators $\big(P_s\,, T^i\,, \break T^{(m_1 \cdots m_n)k} \big)$
can be called ${\cal K}$. By analogy with the interpretation of the Einstein gravity as a nonlinear realization of the group {\it Diff}$\,R^4$
it was suggested in \cite{IvOg1} that the Yang-Mills theory can be interpreted as the appropriate nonlinear realization of ${\cal K}$.
An essential difference from the case of gravity is that ${\cal K}$ cannot be obtained as a closure of any two finite-dimensional
subgroups. Actually, the only closed non-trivial subgroup of ${\cal K}$ is the original internal symmetry group $G_0$ generated by $T^i$ subjected to
\p{Global}. So the appearance of an infinite number of Goldstone fields in any nonlinear realization of ${\cal K}$ seems inevitable. Fortunately,
most of such fields are eliminated by the inverse Higgs effect.

Thus, let us consider the realization of ${\cal K}$ by the left shifts on the coset manifold ${\cal K}/G_0$. The coset element can be
written as
\bea
G(x, b) = {\rm e}^{ix^mP_m}{\rm e}^{i\sum_{n\geq 1}\frac1{n!}\,b^i_{m_1\cdots m_n}(x) T^{(m_1 \cdots m_n)i}}\lb{CosYM}
\eea
and under the left ${\cal K}$ multiplications is transformed as
\bea
G(x, b') = g G(x, b){\rm e}^{-iu^i(x,b, g)T^i}\,, \quad g = {\rm e}^{ia^kT^k}\,
{\rm e}^{i\sum_{n\geq 1}\frac1{n!}\,a^k_{m_1\cdots m_n}T^{(m_1 \cdots m_n)k}}\,.\lb{Defg}
\eea
The `matter'' fields $\Psi^\alpha$, in accord with the general rules of nonlinear realizations \cite{nonl12}, are transformed as
\be
\Psi^{\alpha}{}' = ({\rm e}^{iu^i(x,b, g)\hat{T}^i})^{\alpha}_{\beta}\Psi^{\beta}\,,\lb{otherTrans}
\ee
where $\hat{T}^i$ are the  matrix realization of the generators $T^i$ in the representation of $G_0$ by which $\Psi^{\alpha}$ is  transformed.

The first factor in $g$ \p{Defg} just homogeneously rotates all fields with respect to the adjoint representation index $i$, so $u^i(x,b, g) = a^i$
in this case and
\p{otherTrans} yields global $G_0$ transformation of $\Psi^{\alpha}$. The parameters of the second factor generate some nonlinear
inhomogeneous transformations
of the coset fields like $\delta b^i_{m_1\cdots m_n}(x) =
a^i_{m_1\cdots m_n} + O(b)$. Using the commutation relations \p{PTm}, \p{InfDim} it is rather direct to establish that
$u^k(x,b, g) = \sum_{n\geq 1}\frac1{n!}\,a^k_{m_1\cdots m_n}x^{m_1}\ldots x^{m_n} := \lambda^k(x)$ in this case. In other words,
the induced $G_0$ transformation
is just the standard $G_0$ gauge transformation. Well, where is then the gauge field? To answer this question, we need to construct
the corresponding Cartan forms
and the covariant derivative of $\Psi^\alpha$
\bea
&& G^{-1}d G = idx^mP_m + i{\cal V}^k_mdx^m T^k + i\sum_{n\geq 1}\frac1{n!}\,\nabla_s b_{(m_1\cdots m_n)}^k dx^s T^{(m_1 \cdots m_n)k}\,,
\lb{CartYM} \\
&& {\cal D}_m\Psi^\alpha = \partial_m\Psi^\alpha + i{\cal V}^k_m (\hat{T}^k)^\alpha_{\;\;\beta}\Psi^\beta\,. \lb{DerYM}
\eea
It is easy to compute
\bea
&& {\cal V}^k_m = b_m^k\,, \qquad \nabla_m b_n^i = \partial_m b^i_n + b^i_{(mn)} - \frac12 c^{ikl}b^k_m b^l_n\,, \lb{CovYMDer}\\
&& \nabla_s b_{(m_1\cdots m_n)}^i = \partial_s b_{(m_1\cdots m_n)}^i + \ldots \,,  \lb{CovYMDer1}\\
&& \delta b_m^i = -\partial_m \lambda^i + c^{ikl}b^k_m\lambda^l\,, \quad \delta \nabla_m b_n^i =  c^{ikl}(\nabla_m b_n^k) \lambda^l\,,
\;\; {\rm etc}. \lb{TrYM}
\eea
{}From \p{DerYM}, \p{CovYMDer} and \p{TrYM} we observe that $b^i_m$ possesses the standard transformation properties of the Yang-Mills field and
enters the covariant derivative of $\Psi^\alpha$ in the right way. Further, we observe that the skew-symmetric and symmetric parts
of the covariant derivative
$\nabla_m b_n^i$,
\be
2\nabla_{[m} b_{n]}^i = \partial_m b^i_n -  \partial_n b^i_m - c^{ikl}b^k_m b^l_n\,, \quad 2\nabla_{(m} b_{n)}^i = \partial_m b^i_n +
\partial_n b^i_m + 2b^i_{(mn)}\,,
\ee
are covariant separately. The skew-symmetric part is just the covariant field strength of $b_m^i$, while the symmetric
part can be put equal to zero,
yielding the inverse Higgs expression for $b^i_{mn}$
\be
\nabla_{(m} b_{n)}^i = 0 \quad \Rightarrow \quad b^i_{(mn)} = - \partial_{(m} b^i_{n)}\,. \lb{1stIHE}
\ee
As follows from \p{PTm}, \p{InfDim}, the commutators of $P_m$ with all spontaneously broken generators $T^{(m_1 \cdots m_n)i}$
besides $T^{m\, i}$ contains
the generators of the same (spontaneously broken) type, so all the related Goldstone fields  can be eliminated by the inverse Higgs effect. This can be accomplished,
like in \p{1stIHE},
by equating to zero the totally symmetric parts of the appropriate covariant derivatives
\be
\nabla_{(s} b_{m_1 \cdots m_n)}^i = 0\,.\lb{IHgauge}
\ee
The commutator \p{PTm} yields the generator of the stability subgroup, so the Goldstone (gauge) field $b_m^i$ cannot be eliminated, and it is the only ``true''
Goldstone field in the considered case.

In the paper \cite{YMsigma} a general solution of \p{IHgauge} was found. The abstract algebra ${\cal K}$ was realized as
\footnote{Our conventions here are slightly
different from those in \cite{YMsigma}.}
\be
P_m = \frac1{i} \frac{\partial}{\partial y^m}\,, \quad T^{(m_1 \cdots m_n)i} = y^{m_1}\ldots y^{m_n} T^i\,,
\ee
where $y^m$ is some new 4-vector coordinate. Then the coset element \p{CosYM} can be rewritten in the concise form as
\be
G(x,y) = {\rm e}^{ix^mP_m}{\rm e}^{i b^k(x, y)T^k}\,, \quad b^k(x, y) =
\sum_{n\geq 1} \frac1{n!}b_{(m_1\ldots m_n)}^k(x)\,y^{m_1}\ldots y^{m_n}\,, \quad b^k(x, 0)= 0\,,
\ee
while the covariant derivatives of the Goldstone fields corresponding to the Cartan forms \p{DerYM} as
\bea
&&{\rm e}^{-i b^k(x, y)T^k}(\partial_m^{x} + \partial_m^{y}){\rm e}^{-i b^k(x, y)T^k} = i\omega_m^k(x, y)T^k\,, \nn \\
&& \omega_m^k(x, y) = \,b_m^k(x)  + \sum_{n\geq 1} \frac1{n!}\nabla_m b_{(m_1\ldots m_n)}^k(x)\,y^{m_1}\ldots y^{m_n}\,.
\eea
The inverse Higgs constraints \p{IHgauge} in this formalism are rewritten as
\be
y^m(\partial_m^x + \partial^{y}_m){\rm e}^{-i b^k(x, y)T^k} = -iy^mb^k_m(x)T^k\,{\rm e}^{-i b^k(x, y)T^k} \lb{IHguge1}
\ee
and are solved by
\be
{\rm e}^{-i b^k(x, y)T^k} = P \exp \{i \int^x_{x - y} b_m^k(\xi)T^k d\xi^m\}\,,
\ee
where $P$ denotes ordering in the matrices $T^i$ along the straight line connecting the points $(x-y)^m$ and $x^m$. This representation
could be suggestive for identifying the hypothetical {\it linear} sigma model for gauge fields corresponding to the nonlinear realization
constructed above.

To summarize, all gauge theories including gravity and Yang-Mills theory, correspond to the spontaneous breaking of some underlying symmetry,
finite- or infinite-dimensional, and can be consistently derived by applying the general nonlinear realizations machinery
to these symmetries. The inverse Higgs phenomenon plays a crucial role in this derivation. \\

\noindent{\bf 2.8 Gravitation theories as gauge theories.} Finally, it is worth to mention here the papers \cite{INie1}, \cite{INie2}
closely related to the circle of problems discussed in this Section.

Actually, for a long time there were certain difficulties with the treatment of gravitation theories as gauge theories. The most direct analogy
with the Yang-Mills theories is achieved, when treating gravity as a gauge theory associated with local symmetries
in the tangent space, rather than with {\it Diff}$\,R^4$ obtained by gauging rigid space-time symmetries, like $x^m$-translations and Lorentz rotations.
The basic objects in such formulations are the direct and inverse vierbeins $e_m^a$ and $e^{ma}$ treated as gauge fields for some translation-like
generators in the tangent space.  The basic problem with such formulations was how to covariantly eliminate other
gauge fields associated with these tangent-space groups, which include some other generators in parallel with the translation-like ones.
In our papers \cite{INie1,INie2} with Jiri Niederle (1939 - 2010), the correct way of deriving various versions of gravitation
theories by gauging the groups
in the tangent space was formulated. It was shown there how to construct the correct formulations
which make manifest the deep analogies of the gravitation theories with the standard Yang-Mills theories and ensure the covariance
of the conditions eliminating the redundant gauge fields (actually, these constraints are quite similar to the inverse Higgs conditions). One
should treat the tangent
space gauge groups as spontaneously broken ones, with some additional Goldstone fields associated to the translation-like
generators\footnote{A similar proposal was made in \cite{SW1}.}. Then the vierbeins are to be  identified
with the covariant derivatives of such fields, rather than directly with
the gauge fields (e.g., in the Einstein gravity, $e_m^a = \partial_m \phi^a + \ldots$, where $\phi^a(x)$
is the Goldstone field parametrizing the spontaneously broken tangent space 4-translations). The standard gravity theories naturally come out
after choosing the ``soldering'' gauge, in which these extra Goldstone fields are
identified with the space-time coordinates. It was also explicitly shown that the diverse gravity theories which differ in the maximally
symmetric classical backgrounds
(e.g., Poincar\`e,  de Sitter and anti-de Sitter gravities, Weyl gravity, etc) correspond to gauging different tangent space groups
having as the common important feature
the presence of some spontaneously broken translation-like generators (in general, corresponding to curved ``translations''). \\
\setcounter{equation}{0}

\section{Supersymmetry: Early years}
%%%%%%%%%%%%%%%%%%%%%%%%%%%%%%%%%%%%%%%%%%%%%%%%%%%%%%%%%%%%%%%%%%%%

The invention of supersymmetry at the beginning of the seventies \cite{susy}-\cite{susy2} sharply influenced
the further fate of the mainstream research activity in the Markov Group. V.I. Ogievetsky rapidly realized the
potential importance of this new concept for the particle and mathematical theoretical physics. One of the discoverers
of supersymmetry was Dmitry Vasil'evich Volkov (1925 - 1996) from the Kharkov Institute of Physics and Technology.
For a long time he and his group had close scientific and
human contacts with Ogievetsky and his collaborators. So it is not surprising that just the study of supersymmetric theories
became the basic research direction in the group of young researchers formed and headed by Ogievetsky.
The active members of this team were Luca Mezincescu from Bucharest and Emery Sokatchev
from Sofia. Boris Zupnik, who defended his PhD to that time and received a position in the Institute of Nuclear Physics
in Ulugbek near Tashkent, has  penetrated deeply into this new area despite a few thousand  kilometers separating Dubna and Ulugbek (Boris
succeeded to come back to Dubna and join our group only in 1994).
The author, after defending his PhD in 1976 under supervision of V.I. Ogievetsky, has  also focused on this line of  investigations.
Later on, this research team constituted the main staff of sector ``Supersymmetry'' in BLTP headed by V.I. Ogievetsky.
In the beginning of eighties it was enriched by the talented and enthusiastic PhD student Sasha Galperin from Tashkent.
Also, I would like to mention Stilian Kalitzin from  Sofia. Like A. Galperin, he was a representative of the second generation
of the supersymmetry adepts. To the same category I would
attribute Alexander Sorin, my first PhD student. The permanent contacts with Volkov group, as well as with the Fradkin group
from Lebedev Institute, were certainly very helpful and conducive for the successful development of  investigations
on supersymmetry in Dubna. In particular, an essential contribution to this development was brought by Volkov's scholar
Anatoly Pashnev (1948 - 2004) who was employed  on the contract in BLTP from Kharkov in 1991 and worked in Dubna until his untimely and tragic death.
Frequent visits to Dubna by Sasha Kapustnikov (1945 - 2003) from Dniepropetrovsk, my friend and co-author for many years,
have also played an invaluable role. More recently, in the nineties, there were established  the fruitful and firm contacts
with the research groups of the theorists from Tomsk,
Joseph Buchbinder and Anton Galajinsky.

Since the start of the supersymmetry epoch, the basic interests of Ogievetsky and his surrounding were concentrated
on the superspace approach to supersymmetric theories.\\

\noindent{\bf 3.1 Invariant actions in superspace. }
The natural arena for supersymmetry is superspace, an extension of some bosonic space by anticommuting
fermionic (Grassmann) coordinates. For the ${\cal N}=1$ Poincar\'e supersymmetry
it was introduced in \cite{VA} as a
coset of the ${\cal N}=1$ Poincar\'e supergroup over its bosonic Lorentz subgroup.
However, the fermionic coset parameters were treated in \cite{VA} as Nambu-Goldstone fields
``living'' on Minkowski space and supporting a nonlinear realization of ${\cal N}=1$ Poincar\'e supersymmetry.
It was suggested by Salam and Strathdee \cite{SS} to treat the fermionic coordinates $\theta^\alpha\,, \bar\theta^{\dot\alpha}\,, (\alpha, \dot\alpha = 1,2)\,,$ on
equal footing with $x^m$ as {\it independent} coordinates.
The fields on such an extended space (superspace) were christened  superfields. They naturally encompass the irreducible
${\cal N}=1$ supermultiplets the fields of which appear as coefficients in the expansions of superfields over Grassmann
coordinates. The remarkable property of superfields is that these expansions terminate at a finite step due to the nilpotency
of the Grassmann coordinates \cite{SS,Ferr}. Another advantage of superfields is the simple rule of constructing component actions
invariant under supersymmetry. In any products of superfields and their ordinary and/or covariant spinor derivatives
the highest components in the expansions of these products  over Grassmann coordinates (the so-called $D$ component, if the product
is a general superfield, and the $F$ component, if the product is a chiral superfield) is transformed to a total $x$-derivative and
so is a candidate for the supersymmetric action in the Minkowski space-time.

In \cite{OM1}, Ogievetsky and
Mezincescu proposed an elegant way of writing down the invariant superfield
actions directly in superspace. As just mentioned, the invariant actions can be constructed as the
$x$-integrals of the coefficients of the highest-degree $\theta$ monomials in
the appropriate products of the involved superfields. The question was how to
extract these components in a manifestly supersymmetric way. Ogievetsky and
Mezincescu proposed to use the important notion of Berezin integral \cite{Ber}
for this purpose. In fact, Berezin integration is equivalent to the Grassmann
differentiation and, in the case of ${\cal N}=1$ superspace, is defined by the rules
\be
\int d\theta_\alpha\, \theta^\beta = \delta^\beta_\alpha\,, \quad \int
d\theta^\alpha\, 1 = 0\,, \quad \{d\theta_\alpha, d\theta_\beta \}
=\{\theta_\alpha, d\theta_\beta \} = 0\,, \lb{BerInt}
\ee
and analogous ones for the conjugated coordinates $\bar\theta^{\dot\alpha}$.
It is easy to see
that, up to the appropriate normalization,
\be
\int d^2 \theta \,(\theta)^2 =
1\,, \;\; \int d^2 \bar\theta\, (\bar\theta)^2 = 1\,, \;\; \int d^2\theta
d^2\bar\theta\, (\theta)^4 = 1\,, \lb{BerInt2}
\ee
and, hence, Berezin
integration provides the manifestly supersymmetric way of extracting
the coefficients of the highest-order $\theta$ monomials. For
example, the simplest invariant action of chiral superfields can be written as
\be
S \sim \int d^4x d^4\theta\, \varphi(x_L, \theta)\bar\varphi (x_R,
\bar\theta)\,, \quad x_L^m = x^m + i\theta \sigma^m\bar\theta\,, \; x^m_R = \overline{(x_L^m)}
\,, \lb{ChirAct}
\ee
where the superfields satisfy the chirality and anti-chirality conditions
\be
\bar{D}_{\dot\alpha}\varphi(x_L, \theta) = 0\,, \quad {D}_{\alpha}\bar\varphi(x_R, \bar\theta) = 0\,, \lb{Chirality}
\ee
with
\be
D_\alpha =\frac{\partial}{\partial \theta^\alpha} +
i(\sigma^m\bar\theta)_\alpha \partial_m\,, \; \bar D_{\dot\alpha} =
-\frac{\partial}{\partial \bar\theta^{\dot\alpha}}-
i(\theta\sigma^m)_{\dot\alpha} \partial_m\,,\; \{D_\alpha, \bar
D_{\dot\alpha}\}= -2i (\sigma^m)_{\alpha\dot\alpha}\partial_m\,.  \lb{DerivSP}
\ee

Using the $\theta$ expansion of the chiral superfield $\varphi(x_L, \theta)$,
\be
\varphi(x_L, \theta) = \varphi(x_L) + \theta^\alpha\psi_\alpha(x_L) + (\theta)^2 F(x_L)\,,
\ee
and its conjugate $\bar\varphi(x_R, \bar\theta)$, it is easy to
integrate over $\theta, \bar\theta$ in \p{ChirAct} and, discarding total
$x$-derivatives,  to obtain the component form of the action
\be
S \sim \int
d^4 x \left(\partial^m\bar\varphi \partial_m\varphi -\sfrac{i}{2}
\psi\sigma^m\partial_m\bar\psi + F\bar F \right).\lb{FreeWZ}
\ee
It is just the
free action of the massless scalar ${\cal N}=1$ multiplet. It can be generalized
to the case with interaction, choosing the Lagrangian as an arbitrary
function $K(\bar\varphi, \varphi)$ and adding independent potential terms
\be
\sim \int d^4x_L d^2\theta\, P(\varphi) + \mbox{c.c.}\,. \lb{PoT}
\ee
The sum of \p{ChirAct} and the superpotential term \p{PoT} with $P(\varphi) \sim g\varphi^3 + m
\varphi^2$ yields the Wess-Zumino model \cite{WZu} which was the first
example of nontrivial ${\cal N}=1$ supersymmetric model and the only renormalizable
model of scalar ${\cal N}=1$ multiplet. Ogievetsky and Mezincescu  argued
that the representation of the action of the Wess-Zumino model in
terms of Berezin integral is very useful and suggestive, while developing the
superfield perturbation theory for it. All quantum
corrections have the form of the integral over the whole ${\cal N}=1$ superspace, so
the superpotential term (and, hence, the parameters $g$ and $m$) is not
renormalized. This was the first example of the non-renormalization theorems, which
nowadays are the powerful ingredients of the quantum superfield approach.

In 1975, Ogievetsky and Mezincescu wrote a comprehensive review on the basics of supersymmetry and
superspace techniques \cite{Rev}. Until present it is still one of the best introductory reviews in
this area.\\

\noindent{\bf 3.2 Superfields with definite superspins and supercurrents.}
The dream of Ogievetsky was to generalize the spin principle formulated by him and Polubarinov to the
superfield approach. Indeed, the notion of the Poincar\'e spin of fields naturally extends
to the case of supersymmetry as the notion of superspin, the eigenvalue of one of the Casimir operators of
the Poincar\'e supersymmetry algebra.  The irreducible ${\cal N}=1$  supermultiplets
are characterized by definite superspins, and the latter can be well defined  for the interacting
superfields, like spin in the Poincar\' e invariant theories. The means to ensure the definite spin
of superfields are either the appropriate irreducibility constraints (in the massive case) or the appropriate
gauge invariance (in the massless case). The concept of conserved current also admits
``supersymmetrization'', and the appropriate supercurrents were already  known for a number of simple models.
The requirement of preserving definite superspins by interacting superfields was expected to fully determine the structure of the corresponding
actions, as well as the gauge group intended to make harmless extra superspins carried
by the given off-shell superfield. However, because of existence of new differential operators in the superfield case,
the covariant spinor derivatives \p{DerivSP}, along with the standard $x$-derivative,
yet defining the irreducibility  conditions for most cases of interest
was very difficult technical problem.  In the pioneering paper \cite{SS}, the decomposition into the superspin-irreducible parts
was discussed only for a scalar ${\cal N}=1$ superfield.

The general classification of ${\cal N}=1$ superfields by superspin was
given by Sokatchev in the paper \cite{ES1}, where the corresponding irreducibility superfield constraints,
together with the relevant projection operators on definite superspins, were found.
In fulfilling the program of generalizing the spin principle to supersymmetry, the
formalism of the projection operators of \cite{ES1} proved to be of key significance.

The main efforts of Ogievetsky and Sokatchev were soon concentrated on seeking
a self-consistent theory of massless axial-vector superfield (carrying superspins 3/2 and 1/2).
This superfield $H^{n}(x,\theta, \bar\theta)$ was of special interest because its
component field expansion involved a massless tensor field $e_{a}^n$ and the spin-vector field $\psi_{\alpha}^n\,$,
$$
H^n = \theta\sigma^a\bar\theta e_{a}^n + (\bar\theta)^2 \theta^\alpha \psi_{\alpha}^n +
(\theta)^2 \bar\theta_{\dot\alpha}\bar\psi^{\dot\alpha n} + \ldots\,.
$$
These fields could naturally be identified with those of graviton and gravitino of ${\cal N}=1$
supergravity (SG) known to that time in the component form \cite{SG}. In
\cite{ES3}, Ogievetsky and Sokatchev have put forward the hypothesis that the correct
``minimal'' ${\cal N}=1$ superfield SG should be a theory of gauge
axial-vector superfield $H^m(x,\theta, \bar\theta)$ generated  by the conserved
supercurrent. The latter unifies into an irreducible ${\cal N}=1$ supermultiplet the
energy-momentum tensor and spin-vector current associated with the
supertranslations \cite{FerrZu} (see also \cite{SCurr}, \cite{SEM} and refs. therein). Ogievetsky
and Sokatchev relied upon the clear analogy with the Einstein gravity which can
be viewed as a theory of massless tensor field generated by the conserved
energy-momentum tensor. As was already mentioned in Sect. 2, the whole Einstein action and its non-abelian $4D$
diffeomorphism gauge symmetry can be uniquely restored step-by-step, starting
with a free action of symmetric tensor field and requiring its source
(constructed from this field and its derivatives, as well as from matter
fields) to be conserved \cite{OP2}. In \cite{ES3}, this Noether procedure was
applied to the free action of $H^m(x,\theta, \bar\theta)\,$. The first-order
coupling of $H^m$ to the conserved supercurrent of the matter chiral superfield
was restored and the superfield gauge symmetry generalizing bosonic diffeomorphism
symmetry was identified at the linearized level. The geometric meaning of this
supergauge symmetry and its full non-abelian form were revealed by Ogievetsky
and Sokatchev later, in the remarkable papers \cite{ESsg,ESsg1}. \\

\noindent{\bf 3.3 Complex superfield geometry of ${\cal N}=1$ supergravity.}
After the discovery of the component ${\cal N}=1$ supergravity in \cite{SG} \footnote{A version of supergravity
with the spontaneously broken supersymmetry
(based on the Higgs effect for the Goldstone fields with the spin $1/2$) was worked out in \cite{VSa} (see also a recent paper \cite{BandS}).},  it
was an urgent problem to find its complete off-shell formulation, {\it i.e.},
to extend the set of physical fields of graviton and gravitino to an off-shell multiplet by
adding the appropriate auxiliary fields and/or to formulate ${\cal N}=1$ supergravity in
superspace, making all its symmetries manifest.

One of the approaches to the superspace formulations of ${\cal N}=1$ supergravity was to start
from the most general differential geometry in ${\cal N}=1$ superspace. One defines supervielbeins, supercurvatures
and supertorsions which are covariant under arbitrary ${\cal N}=1$ superdiffeomorphisms, and then imposes
the appropriate covariant constraints, so as to single out the minimal set of off-shell ${\cal N}=1$ superfields
carrying the irreducible field content of supergravity \cite{W}. An alternative approach would consist
of revealing the fundamental minimal gauge group of supergravity and defining the basic unconstrained
prepotential, an analog of ${\cal N}=1$ super Yang-Mills (SYM) prepotential \cite{SYM}. This was just the strategy which Ogievetsky
and Sokatchev follow in \cite{ESsg} to construct a beautiful geometric formulation of
the conformal and ``minimal'' Einstein ${\cal N}=1$ SG \footnote{A closely similar formulation was worked out in the parallel investigations
by Siegel and Gates \cite{SiGa}.}.

It is based on a generalization of the notion of ${\cal N}=1$ chirality to the
curved case. The flat chiral ${\cal N}=1$ superspace $(x_L^m, \theta^\mu_L)$ possesses
the complex dimension $(4|2)$ and contains the ${\cal N}=1$ superspace $(x^m,
\theta^\mu, \bar\theta^{\dot\mu})$ as a real $(4|4)$ dimensional hypersurface
defined by the following embedding conditions
\bea
\mbox{(a)} \;\; x^m_L +
x^m_R = 2 x^m\,, \quad \mbox{(b)}\;\;  x^m_L - x^m_R =
2i\theta\sigma^m\bar\theta\,, \quad \theta^\mu_L = \theta^\mu\,, \;
\bar\theta^{\dot\mu}_R = \bar\theta^{\dot\mu}\,, \label{Emb}
\eea
and $x^m_R = \overline{(x^m_L)}, \bar\theta^{\dot\mu}_R = \overline{(\theta_L^\mu)}\,$.
The underlying gauge group of conformal ${\cal N}=1$ SG proved to be the
group of general diffeomorphisms of the chiral superspace:
\be
\delta x^m_L =
\lambda^m(x_L, \theta_L)\,, \quad \delta \theta^\mu_L = \lambda^\mu(x_L,
\theta_L)\,,\lb{ChirDif}
\ee
where $\lambda^m, \lambda^\mu$ are arbitrary
complex functions of their arguments. The fermionic part of the embedding
conditions \p{Emb} does not change, while the bosonic part is generalized as
\be
\mbox{(a)} \;\;x^m_L + x^m_R = 2 x^m\,, \quad \mbox{(b)}\;\;  x^m_L - x^m_R
= 2i H^m(x, \theta, \bar\theta)\,. \lb{Emb1}
\ee
The basic gauge prepotential of conformal ${\cal N}=1$ SG is just the axial-vector superfield $H^m(x,\theta,
\bar\theta)$ in \p{Emb1}. It specifies the superembedding of real ${\cal N}=1$
superspace as a hypersurface into the complex chiral ${\cal N}=1$ superspace $(x_L^m,
\theta^\mu_L)\,.$ Through relations
\p{Emb1}, the transformations \p{ChirDif} generate field-dependent nonlinear
transformations of the \break ${\cal N}=1$ superspace coordinates $(x^m, \theta^\mu,
\bar\theta^{\dot\mu})$ and of the superfield $H^m(x, \theta, \bar\theta)\,$.
The irreducible field content of $H^m$ is revealed in the WZ gauge which requires
knowing only the linearized form of the transformations:
\bea
\delta_{lin} H^m &=&
\sfrac{1}{2i}\left[\lambda^m(x + i\theta\sigma\bar\theta, \theta) -
\bar\lambda^m(x - i\theta\sigma\bar\theta, \bar\theta)\right] \nn \\
&& -\, \lambda(x + i\theta\sigma\bar\theta, \theta)\sigma^m\bar\theta -
\theta\sigma^m \bar\lambda(x - i\theta\sigma\bar\theta, \bar\theta)\,.\lb{WZH1}
\eea
Here we assumed the presence of the ``flat'' part
$\theta\sigma^m\bar\theta$ in $H^m = \theta\sigma^a\bar\theta(\delta^m_a +
\kappa h^m_a) + \ldots \,$. The WZ gauge form of
$H^m$ is then as follows
\be
H^m_{WZ} = \theta\sigma^a\bar\theta\,e^m_a + (\bar\theta)^2
\theta^\mu \psi^m_\mu + (\theta)^2 \bar\theta_{\dot\mu}\bar\psi^{m\dot\mu} +
(\theta)^2(\bar\theta)^2 A^m\,. \label{WZH}
\ee
Here, one finds the inverse vierbein $e^m_a$ presenting the conformal graviton (gauge-independent spin 2 off-shell),
the gravitino $\psi^m_\mu$ (spin $(3/2)^2$), and the gauge field $A^m$ (spin 1)
of the local $\gamma_5$ R-symmetry. They constitute just $(8 +8)$ off-shell degrees of freedom
of the superspin 3/2 ${\cal N}=1$ Weyl multiplet.

The Einstein ${\cal N}=1$ SG can now be deduced in the two basically equivalent ways.
The first one was used in the
original paper \cite{ESsg} and it consists in restricting the group \p{ChirDif} by the constraint
\be
\partial_m\lambda^m - \partial_\mu \lambda^\mu = 0\,,\lb{Ber1}
\ee
which is the infinitesimal form of the requirement that the integration
measure of chiral superspace $(x_L, \theta^\mu)$ is invariant. One can show
that, with this constraint, the WZ form of $H^m$ collects two extra scalar
auxiliary fields, while $A^m$ ceases to be gauge and also becomes an auxiliary
field. On top of this, there disappears one fermionic gauge invariance
(corresponding to conformal supersymmetry) and, as a result, spin-vector field starts to
comprise  12 independent components. So, one ends up with the (12 + 12) off-shell
multiplet of the so-called ``minimal'' Einstein SG \cite{AuxSG}.

Another, more suggestive way to come to the same off-shell content is to use
the compensator techniques which can be traced back to the interpretation of
Einstein gravity as conformal gravity with the compensating (Goldstone) scalar
field \cite{FT}. Since the group \p{ChirDif} preserves the chiral superspace,
in the local case one can still define a chiral superfield $\Phi(x_L, \theta)$
as an unconstrained function on this superspace and ascribe to it the following
transformation law
\be
\delta \Phi = -\sfrac{1}{3}\left(\partial_m\lambda^m -
\partial_\mu \lambda^\mu\right) \Phi\,,\lb{TranphiL}
\ee
where the specific
choice (-1/3) of the conformal weight of $\Phi$ is needed for constructing the
invariant SG action. One can show that such a compensating chiral superfield
together with the prepotential $H^m$ yield, in the appropriate WZ gauges,  just
the required off-shell $(12 + 12)$ representation.

The basic advantage of the compensating method is the possibility to easily write the action of the minimal
Einstein SG as an invariant action of the compensator $\Phi$ in the background of the Weyl multiplet carried
by $H^m$:
\bea
S_{SG} = - \sfrac{1}{\kappa^2}\int d^4x d^2\theta d^2\bar\theta\, E \,
\Phi(x_L, \theta)\bar\Phi(x_R, \bar\theta)
+ \xi \left(\int d^4x_L d^2\theta \Phi^3(x_L, \theta) + \mbox{c.c.} \right). \lb{SGact}
\eea
Here $E$ is a density constructed from $H^m$ and its derivatives \cite{ESsg1}, such that
its transformation cancels the total weight transformation of the integration measure
$d^4xd^2\theta d^2\bar\theta$ and the product of chiral compensators. In components,
the first term in \p{SGact} yields the minimal Einstein
${\cal N}=1$ SG action without cosmological term, while the second term in \p{SGact} is the superfield form of
the cosmological term $\sim \xi\,$.

Later on, many other off-shell component and superfield versions of ${\cal N}=1$ SG were constructed.
They mainly differ in the choice of the compensating supermultiplet. The uncertainty in choosing compensating superfields
is related to the fact that the same on-shell scalar ${\cal N}=1$ multiplet admits variant off-shell
representations.

The Ogievetsky-Sokatchev formulation of ${\cal N}=1$ SG was one of the main
indications that the notion of chiral superfields and chiral superspace play
the pivotal role in ${\cal N}=1$ supersymmetry. Later it was found that the superfield
constraints of ${\cal N}=1$ SG have the nice geometric meaning: they guarantee the
existence of chiral ${\cal N}=1$ superfields in the curved case, once again pointing
out the fundamental role of chirality in ${\cal N}=1$ theories. The constraints
defining the ${\cal N}=1$ SYM theory can also be derived from requiring chiral
representations to exist in the full interaction case. The parameters of the
${\cal N}=1$ gauge group are chiral superfields, so this group
manifestly preserves the chirality. The geometric meaning of ${\cal N}=1$ SYM
prepotential $V(x,\theta, \bar\theta)$ was discovered in \cite{Iv}. By analogy
with $H^n(x,\theta, \bar\theta)$, the superfield $V$ specifies a real $(4|4)$
dimensional hypersurface, this time in the product of ${\cal N}=1$ chiral superspace
and the internal coset space $G^c/G$, where $G^c$ is the complexification of the
gauge group $G\,$. At last, chiral superfields provide the most general
description of ${\cal N}=1$ matter since any variant off-shell representation of ${\cal N}=1$
scalar multiplet is related to chiral multiplet via duality transformation.

In parallel with these investigations, in the second half of seventies - the beginning of eighties
two other important themes related to supersymmetry were worked out in Sector 3,  which exerted a sound influence
on further developments in this area.\\

\noindent{\bf 3.4 Relation between linear and nonlinear realizations of supersymmetry.}
One of the first known realizations of ${\cal N}=1$ supersymmetry was its nonlinear
(Volkov-Akulov) realization \cite{VA}
\be
y^m{\,}' = y^m + i[\lambda(y) \sigma^m\bar\epsilon - \epsilon\sigma^m\bar\lambda(y)]\,,
\quad \lambda^\alpha{\,}'(y{\,}') =
\lambda^\alpha(y) + \epsilon^\alpha\,, \;\bar\lambda^{\dot\alpha}{\,}'(y{\,}') =
\bar\lambda^{\dot\alpha}(y) + \bar\epsilon^{\dot\alpha}\,, \label{NL}
\ee
where the corresponding Minkowski space coordinate is denoted by $y^m$ to distinguish it from $x^m$
corresponding to the superspace realization
\be
\theta^\alpha{}' = \theta^\alpha + \epsilon^\alpha\,,  \quad
\bar\theta^{\dot\alpha}{}' = \bar\theta^{\dot\alpha} +
\bar\epsilon^{\dot\alpha}\,, \quad x^m{}' = x^m + i(\theta \sigma^m\bar\epsilon
- \epsilon\sigma^m\bar\theta). \lb{Tran}
\ee
In \p{NL}, \p{Tran}, $\epsilon^\alpha$ and
$\bar\epsilon^{\dot\alpha}$ are the mutually conjugated Grassmann transformation
parameters associated with the ${\cal N}=1$ supertranslation generators $Q_\alpha$ and $\bar Q_{\dot\alpha}$.

The main difference between \p{NL} and \p{Tran} is that \p{NL}
involves the Volkov-Akulov ${\cal N}=1$ Goldstone fermion (goldstino) $\lambda (y)\,$, the characteristic feature of which
is the inhomogeneous transformation law under supertranslations corresponding to the spontaneously
broken supersymmetry. It is a field given on Minkowski space, while $\theta^\alpha$ in \p{Tran} is an independent
Grassmann coordinate, and ${\cal N}=1$ superfields support a linear realization of supersymmetry.
The invariant action of $\lambda, \bar\lambda$ is \cite{VA}:
\be
S_{(\lambda)} = \sfrac{1}{f^2}\int d^4 y \, \det E^a_m\,, \quad E^a_m = \delta^a_m +
i\left(\lambda\sigma^a\partial_m\bar\lambda - \partial_m\lambda\sigma^a\bar\lambda \right),\label{VAact}
\ee
where $f$ is a coupling constant ($[f] = -2\,$).

The natural question was as to what is the precise relation between the nonlinear and
superfield (linear) realizations of the same ${\cal N}=1$ Poinacar\'e supersymmetry. The explicit answer
was for the first time presented in \cite{IKa1}-\cite{IKa3}. There we showed that, given the
Goldstone fermion $\lambda (y)$ with the transformation properties \p{NL}, the
relation between two types of the supersymmetry realizations, \p{Tran} and \p{NL},
is accomplished through the following invertible change of the superspace coordinates:
\be
x^m = y^m + i\left[\theta\sigma^m\bar\lambda(y) -
\lambda(y)\sigma^m\bar\theta\,\right], \quad \theta^\alpha =
\tilde{\theta}^\alpha + \lambda^\alpha(y)\,, \; \bar\theta^{\dot\alpha} =
\tilde{\bar\theta}^{\dot\alpha} + \bar\lambda^{\dot\alpha}(y)\,, \lb{NLchang}
\ee
where
\be
\tilde{\theta}^\alpha{\,}' = \tilde{\theta}^\alpha\,. \lb{Inert}
\ee
Then the transformations \p{NL} imply for $(x^m, \theta^\alpha,
\bar\theta^{\dot\alpha})$ just the transformations \p{Tran} and, vice-versa,
\p{Tran} imply \p{NL}. Using \p{NLchang}, any linearly transforming superfield
can be put in the new ``splitting'' basis
\be \Phi(x, \theta, \bar\theta) =
\tilde{\Phi}(y, \tilde{\theta}, \tilde{\bar\theta})\,.\lb{SupSpl}
\ee
Since the new spinor coordinate $\tilde{\theta}^\alpha$ is ``inert'' under ${\cal N}=1$ supersymmetry, eq. \p{Inert}, the
components of $\tilde{\Phi}$ transform as ``sigma-fields''\footnote{Below,  $\delta^\star$ stands for the ``active'' variation,
as distinct from other group variations in this Section which are ``passive''.} ,
\be
\delta^\star \phi(y) =
-i[\lambda(y) \sigma^m\bar\epsilon -
\epsilon\sigma^m\bar\lambda(y)]\partial_m\phi(y)\,, \quad \mbox{etc}\,,
\ee
independently of each other, that explains the adjective ``splitting'' for the basis
$(y^m, \tilde{\theta}^\alpha, \tilde{\bar \theta}^{\dot\alpha})$.

As demonstrated in \cite{IKa3}, irrespective of the precise mechanism of
generating goldstino in a theory with the linear realization of spontaneously
broken ${\cal N}=1$ supersymmetry, the corresponding superfield action can be rewritten in the
splitting basis (after performing integration over the inert Grassmann
variables) as
\be
S_{lin} \sim \int d^4 y \det E^a_m \left [ 1 + {\cal
L}(\sigma, \nabla_a\sigma, ...)\right].
\ee
Here ${\cal L}$ is a function of
the ``sigma'' fields and their covariant derivatives $\nabla_a =
E^m_a\partial_m\,$ only, while $\lambda^\alpha(y)$ is related to the goldstino
of the linear realization through a field redefinition. Thus, the Goldstone
fermion is always described by the universal action \p{VAact}, independently of
details of the given dynamical theory with the spontaneous breaking of ${\cal N}=1$
supersymmetry, in the spirit of the general theory of nonlinear realizations.

The transformation \p{NLchang}, \p{SupSpl} can be easily generalized to chiral superfields
and to higher ${\cal N}$. It proved very useful for exhibiting the low-energy structure of
theories with spontaneously broken supersymmetry \cite{FenGold}, as well as in some other problems (see, e.g., \cite{SeKo} and references therein).
It was generalized to the case of local ${\cal N}=1$ supersymmetry in \cite{SaWe,IKa4}. At present, in connection
with some cosmological problems, a great attention is paid to models in which ${\cal N}=1$
supergravity interacts with the matter superfields constructed solely from the Goldstone fermions \cite{NilpSG}\footnote{See also a recent paper \cite{BandS}.}.
The approach based on \p{NLchang}, \p{SupSpl} (and their generalizations to local supersymmetry) is very appropriate for
constructing such multiplets. Indeed, as follows from the transformation law \p{Inert}, the quantities $\tilde{\theta}^\alpha(x, \theta, \bar\theta),
\bar{\tilde{\theta}}^{\dot\alpha}(x, \theta, \bar\theta)$ are ${\cal N}=1$ superfields  properly constrained
because their dependence on the superspace coordinates basically appear through the dependence on $y^m$. Using the definitions
in \p{NLchang} it is easy to deduce
the corresponding superspace constraints \cite{IKa2}:
\bea
D_\beta \tilde{\theta}^\alpha = \delta^\alpha_\beta
+ i(\sigma^m)_{\beta\dot\beta}\bar{\tilde{\theta}}^{\dot\beta}\partial_m\tilde{\theta}^\alpha\,, \quad
\bar D_{\dot\beta} \tilde{\theta}^\alpha =- i(\sigma^m)_{\gamma\dot\beta}{\tilde{\theta}}^{\gamma}\partial_m\tilde{\theta}^\alpha\,,\lb{SPConstr}
\eea
where $D_\alpha, \bar D_{\dot\alpha}$ are defined in \p{DerivSP}, and the analogous ones for $\bar{\tilde{\theta}}^{\dot\beta}\,$.
Thus $\tilde{\theta}^\alpha$ and $\bar{\tilde{\theta}}^{\dot\beta}$ can be
considered as bricks from which more complicated ${\cal N}=1$ superfields as functions of the goldstino field (and its $x$-derivative)
can be assembled.

The constraints \p{SPConstr} look similar to those derived by Samuel and Wess in \cite{SaWe}. The latter are in fact equivalent to \p{SPConstr} and
can be readily derived using a modification of the variable change \p{NLchang}. They follow by starting from the realization
of ${\cal N}=1$ supersymmetry in
the right-handed chiral superspace
\be
\delta x^m_R = - 2i\epsilon \sigma^m \bar\theta\,, \quad \delta \theta^\alpha = \epsilon^\alpha, \;\delta \bar\theta^{\dot\alpha}
= \bar\epsilon^{\dot\alpha}\,,
\ee
and defining the new complex coordinate
\bea
z^m_+ = x^m_R + 2i \chi(z_+) \sigma^m \bar\theta\,, \quad \chi^\alpha(z_+) = \lambda^\alpha(\tilde{z}_+)\,,
\quad z_+^m = \tilde{z}_+^m + i\lambda(\tilde{z}_+)\sigma^m \bar\lambda(\tilde{z}_+),\lb{zDef}
\eea
\bea
\delta z^m_+ = 2i \chi(z_+) \sigma^m \bar\epsilon\,, \; \delta\tilde{z}_+^m = i [\lambda(\tilde{z}_+) \sigma^m \bar\epsilon
- \epsilon \sigma^m \bar\lambda(\tilde{z}_+)]\,, \; \delta\chi^\alpha(z_+) = \epsilon^\alpha, \; \delta\bar{\chi}^{\dot\alpha}(z_+)
= \bar\epsilon^{\dot\alpha}.
\eea
Next,  we define
\be
\tilde{\theta}^\alpha_+ (x, \theta, \bar\theta)  =  \theta^\alpha - \chi^\alpha(z_+)\,,
\ee
and find the following constraints for this complex spinor ${\cal N}=1$ superfield
\be
D_\beta \tilde{\theta}^\alpha_+ = \delta^\alpha_\beta\,, \quad
\bar D_{\dot\beta} \tilde{\theta}^\alpha_+ =
- 2i(\sigma^m)_{\gamma\dot\beta}\tilde{\theta}^{\gamma}_+\partial_m \tilde{\theta}^\alpha_+\,. \lb{SPconstr1}
\ee
These constraints are just those given in \cite{SaWe} (with $ \tilde{\theta}^\alpha_+$ denoted there as $ \Lambda^\alpha\,$).
One can establish the explicit equivalency relation between
$\tilde{\theta}^\alpha$ and $\tilde{\theta}^\alpha_+\,$.

Another possibility, which is also related to the original transformations through an equivalency change
of the goldstino field, is to start from the left-chiral realization
\be
\delta x^m_L=  2i\theta \sigma^m \bar\epsilon\,, \quad \delta \theta^\alpha = \epsilon^\alpha, \;
\delta\bar\theta^{\dot\alpha} = \bar\epsilon^{\dot\alpha},
\ee
and define
\bea
z_- = x^m_L - 2i\theta \sigma^m \bar\omega(z_-)\,, \quad \omega^\alpha(z_-) = \lambda^\alpha(\tilde{z}_-)\,,
\quad z_-^m = \tilde{z}_-^m - i\lambda(\tilde{z}_-)\sigma^m \bar\lambda(\tilde{z}_-),\lb{zDef1}
\eea
\bea
\delta z^m_- = -2i \epsilon\sigma^m \bar{\omega}(z_-), \, \delta\tilde{z}_-^m = i [\lambda(\tilde{z}_-) \sigma^m \bar\epsilon
- \epsilon \sigma^m \bar\lambda(\tilde{z}_-)], \, \delta\omega^\alpha(z_-) = \epsilon^\alpha, \, \delta\bar{\omega}^{\dot\alpha}(z_-)
= \bar\epsilon^{\dot\alpha}.
\eea
The corresponding composite ${\cal N}=1$ superfield,
\be
\tilde{\theta}^\alpha_- (x, \theta, \bar\theta)  =  \theta^\alpha - \omega^\alpha(z_-)\,,
\ee
satisfies the constraints
\be
D_\beta \tilde{\theta}^\alpha_- = \delta^\alpha_\beta +
2i(\sigma^m)_{\beta\dot\beta}\bar{\tilde{\theta}}^{\dot\beta}_-\partial_m\tilde{\theta}^\alpha_-\,, \quad
\bar D_{\dot\beta} \tilde{\theta}^\alpha_- = 0\,.
\ee
So the superfield $\tilde{\theta}^\alpha_- (x, \theta, \bar\theta)$ is chiral, and one can construct the nilpotent chiral scalar superfield
as the bilinear of these ``bricks''
\be
\varphi =\tilde{\theta}^\alpha_-\tilde{\theta}_{- \alpha}\,, \quad \bar D_{\dot\beta}\varphi =0\,, \quad \varphi^2 = 0\,.
\ee
It is worth pointing out that all components of such a nilpotent superfield are model-independent functions  of the goldstino
field $\lambda^\alpha(x), \bar\lambda^{\dot\alpha}(x)$ and its $x$-derivatives only. In the standard description of ${\cal N}=1$ goldstino
through the nilpotent scalar chiral superfield \cite{Roc,SeKo} the latter still includes a scalar auxiliary field
as the independent one. It is eliminated either through its equations of motion, or by imposing additional differential constraints.\\

\noindent{\bf 3.5 Anti-de-Sitter supersymmetry.} Soon after the discovery of the ${\cal N}=1$ Poincar\'e
supersymmetry as a symmetry of theories in the flat Minkowski space treated as a coset of the Poincar\'e group ${\cal P}^4$ over its Lorentz
subgroup, {\it i.e.} ${\cal P}^4/SO(1,3)$, there arose an interest in analogous
supersymmetries preserving non-flat background solutions of Einstein equations.  The renowned manifolds of this kind are
de Sitter and anti-de-Sitter spaces $dS_4 \sim SO(1,4)/SO(1,3)$ and $AdS_4 \sim SO(2,3)/SO(1,3)$. These are solutions of Einstein equations
with a non-zero cosmological constant, respectively positive and negative, so the study of the relevant supersymmetries was expected
to give some hints why this constant is so small (if non-zero). One more source of interest in these ``curved'' supersymmetries
was related to the important
role of the superconformal group $SU(2,2|4)$ involving such supersymmetries as subgroups, along with the flat ${\cal N}=1$ Poincar\'e supersymmetry.
As was already mentioned, various $4D$ supergravities follow from the conformal supergravity through the compensator mechanism.

The anti-de-Sitter supersymmetry is the easiest one to analyze because it is very similar to ${\cal N}=1$ Poincar\'e supersymmetry and
goes over to it in the limit of infinite anti-de-Sitter radius. While the dS$_4$ spinor comprises 8
independent components, no such doubling as compared to the Minkowski space
occurs for AdS$_4$: the AdS$_4$ spinor is the Weyl one with two complex
components, {\it i.e.} the number of supercharges in the AdS supersymmetry is the same as in the ${\cal N}=1$ Poincar\'e one.
A self-consistent superfield formalism for AdS$_4$ supersymmetry was constructed in \cite{IS1,IS2}.

${\cal N}=1$ AdS$_4$ superalgebra is $osp(1|4) \subset su(2,2|1)$, and it is  defined
by the following (anti)commutation relations:
\bea && \{Q_\alpha, \bar
Q_{\dot\alpha}\} = 2(\sigma^m)_{\alpha\dot\alpha}P_m\,, \quad
\{Q_\alpha, Q_{\beta}\} = \mu (\sigma^{mn})_{\alpha\beta}L_{mn}\,, \nn \\
&& [ Q_\alpha, P_m] = \sfrac{\mu}{2}(\sigma_m)_{\alpha\dot\alpha}\bar Q^{\dot\alpha}\,, \quad
[P_m, P_n] = -i \mu^2 L_{mn}\,. \label{ADS4}
\eea
Here $\mu \sim r^{-1}$ is the inverse radius of AdS$_4\,$ and $L_{mn}$
are generators of the Lorentz $SO(1,3)$ subgroup of $SO(2,3)\propto (P_m, L_{mn})\,$. To eqs. \p{ADS4}
one should add complex-conjugate relations and evident commutators with $L_{mn}\,$.
In the limit $\mu \rightarrow 0$ ($r \rightarrow \infty$), \p{ADS4} go over into  the algebra of ${\cal N}=1$ Poincar\'e supersymmetry.

In \cite{IS1,IS2}, we defined the true AdS$_4$ analogs of the general and chiral ${\cal N}=1$
superfields, as well as the vector and spinor covariant derivatives, invariant
superspace integration measures, etc. Having developed the AdS$_4$ superfield
techniques, we constructed the $OSp(1|4)$ invariant actions generalizing the
actions of the Wess-Zumino model and ${\cal N}=1$ super Yang-Mills theory.
For instance, an analog of the free massless
action \p{FreeWZ} of ${\cal N}=1$ scalar multiplet, with the auxiliary fields
eliminated by their equations of motion, reads
\be
S \sim \int d^4 x\,
a^4(x)\left(\nabla^m\bar\varphi \nabla_m\varphi -\sfrac{i}{4}
\psi\sigma^m\nabla_m\bar\psi +\sfrac{i}{4} \nabla_m\psi\sigma^m\bar\psi  +
2\mu^2\,\varphi\bar\varphi  \right). \lb{ADSWZ}
\ee
Here, $a(x) = \sfrac{2}{1 +
\mu^2 x^2}$ is a scalar factor specifying the AdS$_4$ metric in a
conformally-flat parametrization, $ds^2 = a^2(x)\eta_{mn} dx^m d x^n\,$, and
$\nabla_m = a^{-1}\partial_m\,$\footnote{In general, $\nabla_m$ contains a spin connection,
but it drops out from \p{ADSWZ}.}. Taking into account that $\mu^2 =
-\sfrac{1}{12}R$ where $R$ is the scalar curvature of AdS$_4$, this action matches
the standard form of the massless scalar field action in a curved background.

In \cite{IS2}, the vacuum structure of the general massive
AdS$_4$ Wess-Zumino model was studied. This structure proved to be much richer as compared to
the standard ``flat'' Wess-Zumino model due to the presence of the intrinsic
mass parameter $\mu\,$. It was also shown that both the AdS$_4$ massless Wess-Zumino
model and super Yang-Mills theory can be reduced to their flat ${\cal N}=1$ super Minkowski
analogs via some superfield transformation generalizing the Weyl transformation
\be
\varphi(x)  = a^{-1}(x) \tilde\varphi(x), \quad \psi^\alpha(x) =
a^{-3/2}(x)\tilde{\psi}^\alpha(x)\,,\lb{Weyl}
\ee
which reduces \p{ADSWZ} to
\p{FreeWZ}. The existence of the superfield Weyl transformation was an
indication of the superconformal flatness of the AdS$_4$ superspace (although
this property has been proven much later, in \cite{SFlat}).

The simplest supermultiplets of $OSp(1|4)$ derived for the first
time in \cite{IS1} in the superfield approach and the corresponding
projection operators were used in \cite{WS} to give a nice algebraic interpretation
of the superfield constraints of ${\cal N}=1$ supergravity. The interest in $OSp(1|4)$
supersymmetry has especially grown up in recent years in connection with the famous
AdS/CFT correspondence. For instance, the theories invariant under rigid
supersymmetries in various curved manifolds are now under intensive  study (see, e.g. \cite{AdSa}, \cite{AdSb}), and they are
just generalizations of the AdS supersymmetric models the analysis of which was initiated in \cite{IS1,IS2}.
\setcounter{equation}{0}

\section{Harmonic superspace and all that}

After creating the minimal geometric formulation of ${\cal N}=1$ SG described
in Sect. 3, there was posed a natural question as to how it can be generalized to the most
interesting case of extended supergravities and, as a first step, to ${\cal N}=2$
supergravity. To answer this question, it proved necessary to realize what
the correct generalization of ${\cal N}=1$ chirality to ${\cal N}\geq 2$ supersymmetry is and to
invent a new type of superspaces, the harmonic ones.

It was even unclear how to define, in the suggestive geometric way, the appropriate
${\cal N}=2$ analog of the ${\cal N}=1$ SYM prepotential $V(x, \theta, \bar\theta),\;
\delta V = \frac{i}{2} (\Lambda(x_L, \theta) - \bar{\Lambda}(x_R, \bar\theta)) + {\cal O}(V)$.
While the ${\cal N}=1$ SYM constraints are just the integrability conditions for preserving covariant chirality,
\be
\{{\cal D}_\alpha, {\cal D}_\beta \} = 0\,, \quad \{\bar{{\cal D}}_{\dot\alpha}, \bar{{\cal D}}_{\dot\beta} \} = 0\,,
\ee
their ${\cal N}=2$ counterparts read \cite{N2SYM}
\be
\{{\cal D}^{(i}_\alpha, {\cal D}^{k)}_\beta\} = \{\bar{\cal
D}^{(k}_{\dot\alpha}, \bar{\cal D}^{i)}_{\dot\beta}\} = \{{\cal
D}^{(i}_{\alpha}, \bar{\cal D}^{k)}_{\dot\beta}\} = 0\,. \lb{N2symC}
\ee
Here, ${\cal D}_\alpha^i = D^i_\alpha + i {\cal A}^i_\alpha(x, \theta^i, \bar\theta^k)$ and $i, k = 1,2$
are the doublet indices of the automorphism
group $SU(2)_A$ of ${\cal N}=2$ Poincar\'e
superalgebra. Obviously, these constraints cannot be interpreted as the conditions for preserving ${\cal N}=2$ chirality.
Luca Mezincescu solved these
constraints in the abelian case through an unconstrained prepotential
\cite{LMe}. However, the latter has a non-standard dimension -2, and the
corresponding gauge freedom does not admit a geometric interpretation (equally as a reasonable generalization
to the non-abelian case).

There also existed difficulties with an off-shell description of ${\cal N}=2$ hypermultiplet,
the direct analog of ${\cal N}=1$ chiral multiplet. The natural irreducibility constraints on the relevant
superfield $q^{i}(x, \theta^k, \bar\theta^k)$,
\be D^{(i}_\alpha q^{k)} = \bar
D^{(i}_{\dot\alpha} q^{k)} = 0\,,  \lb{HypC}
\ee
are solved by $q^i = f^i + \theta^{i\alpha}\psi_\alpha + \bar\theta^i_{\dot\alpha}\bar\chi^{\dot\alpha} + \ldots\,$,
but simultaneously put the involved fields on their free mass shell. This is a reflection of the ``no-go'' theorem
\cite{nogo} which states that no off-shell representation for hypermultiplet in its
``complex form'' ({\it i.e.} with bosonic fields arranged into $SU(2)$ doublet) can
be achieved with any {\it finite} number of auxiliary fields. No reasonable way to relax \p{HypC} was known.\\

\noindent{\bf 4.1 Way out: Grassmann harmonic analyticity.}
In \cite{GAn} it was observed that extended supersymmetries, besides
the standard chiral superspaces generalizing the ${\cal N}=1$ one, also admit some other
types of the invariant subspaces which were called ``Grassmann-analytic''. Like
chiral superspaces, these analytic  subspaces are revealed by passing to some new
basis in the original general superspace, such that spinor covariant derivatives with
respect to some subset of Grassmann variables become ``short'' in this basis. Then
one can impose {\it Grassmann} Cauchy-Riemann conditions with respect to these
variables. They preserve the full original supersymmetry, but force the relevant {\it analytic}
superfields to depend on a smaller number of Grassmann coordinates (in a deep analogy with
the chirality conditions \p{Chirality}). As a non-trivial example of such
Grassmann analyticity in extended supersymmetries, in \cite{GAn} the existence of a complex ``$O(2)$ analytic subspace''
in ${\cal N}=2, 4D$ superspace was found. Unfortunately, it can be defined only provided that the full automorphism $SU(2)$
symmetry is broken down to $O(2)$. Despite this, it was natural to assume that the Grassmann analyticity of the similar
type could play the fundamental role in extended supersymmetry and provide the correct generalization of ${\cal N}=1$ chirality.
In \cite{Anat} the hypermultiplet constraints \p{HypC} were shown to imply that
different components of the ${\cal N}=2$ superfield $q^i$ ``live'' on different
$O(2)$-analytic subspaces. Since \p{HypC} is $SU(2)$ covariant, it remained to ``$SU(2)$- covariantize'' the $O(2)$ analyticity.

All these problems were solved in the framework of the harmonic superspace \cite{HSS,Greenf,book}.

${\cal N}=2$ harmonic superspace (HSS) is defined as the product
\be
(x^m, \theta_{\alpha\;i},
\bar\theta_{\dot\beta}^k )\otimes S^2\,. \lb{HSS}
\ee
Here, the internal two-sphere $S^2 \sim SU(2)_A/U(1)$ is represented, in
a parametrization-independent way, by the lowest (isospinor) $SU(2)_A$
harmonics
\be
S^2 \in (u^+_i, u^-_k), \quad u^{+i}u_i^- =1, \quad u^{\pm}_i \rightarrow
\mbox{e}^{\pm i\lambda}u^{\pm}_i~.
\ee
It is required that nothing depends on the $U(1)$ phase $\mbox{e}^{i\lambda}$,
so one effectively deals with the 2-sphere $S^2 \sim SU(2)_A/U(1)$.
The superfields given on \p{HSS} (harmonic ${\cal N}=2$ superfields)
are assumed to admit the harmonic expansions on $S^2$,
with the set of all symmetrized products of $u^+_i, u^-_i$
as the basis. Such an expansion is fully specified by the harmonic $U(1)$ charge
of the given superfield\footnote{Another off-shell approach to ${\cal N}=2$ supersymmetric theories is based on the concept
of projective superspace \cite{Project}, an extension of the ordinary ${\cal N}=2$ superspace by a complex $\mathbb{CP}^1$ coordinate.}.

The main advantage of HSS is that it contains an invariant subspace, the
${\cal N}=2$ {\it analytic} HSS, involving only half of the original Grassmann  coordinates
\bea
&& \left(x^m_A, \theta^+_\alpha, \bar\theta^+_{\dot\alpha},
u^{\pm}_{i}\right)  \equiv \left(\zeta^M, u^{\pm}_i  \right)~, \label{22} \\
&& x^m_A = x^m -2i\theta^{(i}\sigma^m\bar\theta^{k)}u^+_iu^-_k~,
\quad \theta^+_\alpha = \theta_{\alpha}^iu^{+}_i~, \;
\bar\theta^+_{\dot\alpha} = \bar\theta_{\dot\alpha}^iu^{+}_i~. \nonumber
\eea
It is just $SU(2)$ covariantization of the $O(2)$ analytic superspace of ref. \cite{GAn}.
It is closed under ${\cal N}=2$ supersymmetry transformations and is real with respect to
the special involution defined as the product of the ordinary complex
conjugation and the antipodal map (Weyl reflection) of $S^2$.

All ${\cal N}=2$ supersymmetric theories have off-shell formulations in terms of unconstrained
superfields defined on \p{22}, the {\it Grassmann analytic}
${\cal N}=2$ superfields. An analytic superfield $\varphi_{\rm an}^{+n}$ with the harmonic $U(1)$ charge $+n$  satisfies the Grassmann harmonic analyticity
constraints
\bea
&& D^+_\alpha \varphi_{\rm an}^{+n} = \bar D^+_{\dot\alpha} \varphi_{\rm an}^{+n} = 0 \quad \Rightarrow \quad \varphi_{\rm an}^{+n}
= \varphi_{\rm an}^{+n}(\zeta, u)\,, \lb{AnalCond}\\
&& D^\pm_\alpha = D^i_\alpha u^\pm_i\,, \quad \bar D^\pm_{\dot\alpha} = \bar D^i_{\dot\alpha} u^\pm_i\,. \lb{ProjDef}
\eea
These constraints are self-consistent just due to the conditions
\be
\{D^+_{\alpha}, D^+_{\beta}\} = \{\bar D^+_{\dot\alpha}, D^+_{\dot\beta}\} = \{D^+_{\alpha}, \bar D^+_{\dot\beta}\}  =0\,, \lb{N2symCflat}
\ee
which are equivalent to the ``flat'' version of \p{N2symC} (these are their projections on $u^+_i$). The solution \p{AnalCond} is obtained
in the analytic basis, where $D^+_\alpha$ and $\bar D^+_{\dot\alpha}$ are reduced to the partial derivatives with respect to
$\theta^{- \alpha}$ and $\bar\theta^{-\dot\alpha}\,$. The opportunity to choose such a basis is just ensured by the integrability conditions \p{N2symCflat}.\\

\noindent{\bf 4.2 ${\cal N}=2$ matter.} In general case the ${\cal N}=2$ matter  is described by $2n$ hypermultiplet
analytic superfields $q^{+}_{a} (\zeta, u)$ ($\overline{(q^{+}_{a})} =
\Omega^{ab}q^{+}_{b}~,\;$ $\Omega^{ab} = -\Omega^{ba}$; $a,b = 1, \dots 2n$ )
with the following off-shell action \cite{CMP}:
\be  \label{33}
S_q = \int du
d\zeta^{(-4)} \left\{q^{+}_{a}D^{++}q^{+a} + L^{+4}(q^+, u^+, u^-) \right\}.
\ee
Here, $du d\zeta^{(-4)}$ is the charged measure of
integration over the analytic superspace \p{22}, $D^{++} =
u^{+\;i}\frac{\partial}{\partial u^{-i}} - 2i \theta^+\sigma^m\bar\theta^+
\frac{\partial}{\partial x^m}$ is the analytic basis form of one of three
harmonic derivatives one can define on $S^{2}$ (it preserves the harmonic Grassmann analyticity) and the indices are raised and
lowered by the $Sp(n)$ totally skew-symmetric tensors $\Omega^{ab},
\Omega_{ab}$, $\Omega^{ab}\Omega_{bc} = \delta^a_c $.
The crucial feature of the general $q^+$ action \p{33} is
an infinite number of auxiliary fields coming from the harmonic expansion on
$S^2$. Just this fundamental property made it possible  to evade the no-go theorem about the non-existence
of off-shell formulations of the ${\cal N}=2$ hypermultiplet in the complex form. The
on-shell constraints \p{HypC} (and their nonlinear generalizations) amount to
{\it both} the harmonic analyticity of $q^{+ a}$ (which is a kinematic property
like ${\cal N}=1$ chirality) and the dynamical equations of motion following from the
action \p{33}. After eliminating infinite sets of auxiliary fields by their
algebraic equations, one ends up with the most general self-interaction of $n$
hypermultiplets. In the bosonic sector it yields the generic sigma model with
$4n$-dimensional hyper-K\"ahler (HK) target manifold, in accord with the theorem
of Alvarez-Gaum\'e and Freedman about the one-to-one correspondence between
${\cal N}=2$ supersymmetric sigma models and HK manifolds \cite{AGF}. In general, the
action \p{33} and the corresponding HK sigma model possess no any isometries.
The object $L^{+4}$ is the HK potential \cite{HK}, an analog of the K\"ahler potential of
${\cal N}=1$ supersymmetric sigma models \cite{Zumino}. Choosing one or another specific $L^{+4}$, one
gets the explicit form of the relevant HK metric by eliminating the auxiliary
fields from \p{33}. So the general hypermultiplet action \p{33} provides an
efficient universal tool of the {\it explicit} construction of the HK metrics \cite{CMP,GIOT}.

The appearance of the HK geometry prepotential as the most general hypermultiplet
interaction superfield Lagrangian is quite similar to the way how the K\"ahler geometry potential
appears as the most general sigma-model super Lagrangian for ${\cal N}=1$ chiral
superfields \cite{Zumino}. In many other cases,  the superfield Lagrangians describing
the sigma-model type interactions of the matter multiplets of diverse supersymmetries prove
also to coincide with the fundamental objects (prepotentials) of the relevant
target complex geometries (see, e.g., \cite{DeldIvNonl} and references therein).
\\

\noindent{\bf 4.3 ${\cal N}=2$ super Yang-Mills theory.} The HSS approach makes manifest that
the ${\cal N}=2$ SYM constraints \p{N2symC} are the integrability conditions for the existence of
the harmonic analytic superfields in such an interacting theory, like in the flat case\footnote{An interpretation of the
constraints of ${\cal N}=2, 3, 4$ SYM theories as the integrability conditions along some directions in the (complexified)
automorphism group manifolds was given by A. Rosly \cite{Rosly}.}. They are
solved in terms of the fundamental geometric object of ${\cal N}=2$ SYM theory, the analytic harmonic connection $V^{++}(\zeta,u)\,$,
which covariantizes the analyticity-preserving harmonic derivative:
\bea
D^{++}
\rightarrow {\cal D}^{++} = D^{++} +i V^{++}~, \quad (V^{++})' = {1\over i}
\mbox{e}^{i\omega}\left( D^{++} + i V^{++} \right)\mbox{e}^{-i\omega}~,
\eea
where
$\omega(\zeta,u)$ is an arbitrary analytic
gauge parameter containing infinitely many component gauge parameters in its
combined $\theta, u$-expansion. The harmonic connection $V^{++}$ contains
infinitely many component fields, however almost all of them can be gauged away
by $\omega(\zeta,u)$. The rest of the $(8+8)$ components is just the off-shell
${\cal N}=2$ vector multiplet. More precisely, in the WZ gauge $V^{++}$ has
the following form:
\bea
V^{++}_{WZ} &=& (\theta^+)^2 w(x_A) + (\bar\theta^+)^2
\bar{w}(x_A) + i\theta^+\sigma^m\bar\theta^+ V_m(x_A)
+ (\bar\theta^+)^2\theta^{+\alpha} \psi_{\alpha}^i(x_A)u^-_i \nn \\
&+& (\theta^+)^2\bar\theta^+_{\dot\alpha}\bar\psi^{\dot\alpha i}(x_A)u^-_i +
(\theta^+)^2(\bar\theta^+)^2 D^{(ij)}(x_A)u^-_iu^-_j ~.
\eea
Here, $V_m, w,\bar
w, \psi^{\alpha}_i, \bar\psi^{\dot\alpha i}, D^{(ij)}$ are the vector gauge field,
complex physical scalar field, doublet of gaugini and the triplet of auxiliary
fields, respectively. All the geometric quantities of ${\cal N}=2$ SYM theory
(spinor and vector connections, covariant superfield strengths, etc.), as well
as the invariant action, can be expressed in terms of
$V^{++}(\zeta,u)$. The closed $V^{++}$ form of the ${\cal N}=2$ SYM
action was found by Boris Zupnik \cite{ZUP1}:
\bea
S_{\rm SYM}^{({\cal N}=2)} =\frac{1}{2g^2}\sum^\infty_{n=2}\frac{(-i)^n}{n}{\rm Tr}\int d^4xd^8\theta du_1\ldots du_n\,
\frac{V^{++}(x,\theta, u_1)\ldots V^{++}(x,\theta, u_n)}
{(u^+_1u^+_2)\ldots (u^+_nu^+_1)}\,,\lb{Zupn}
\eea
where $(u^+_1u^+_2),\ldots\,, (u^+_nu^+_1)$ are the harmonic distributions defined in \cite{Greenf}. An important role is played by the second,
non-analytic gauge connection $V^{--}\,$, which covariantizes the second harmonic derivative $D^{--}$ on the harmonic sphere $S^2$
and is related to $V^{++}$ by
the harmonic flatness condition
\be
D^{++}V^{--} - D^{--}V^{++} + i[V^{++}, V^{--}] = 0\,. \lb{harmflat}
\ee
Most of the objects of the ${\cal N}=2$ SYM differential geometry have a concise representation just in terms of $V^{--}\,$.\\

\noindent{\bf 4.4 ${\cal N}=2$ conformal supergravity.} The ${\cal N}=2$ Weyl multiplet is represented in HSS
by the analytic vielbeins covariantizing $D^{++}$ with respect to
the analyticity-preserving diffeomorphisms of the superspace
$\left(\zeta^M, u^{\pm i}\right)$ \cite{SGconf0,SGconf}:
\bea
&&D^{++} \rightarrow {\cal D}^{++} = u^{+\,i}\frac{\partial}{\partial u^{-i}}
+ H^{++\, M}(\zeta, u)\frac{\partial}{\partial \zeta^M} +
H^{++++}(\zeta, u)u^{-i}\frac{\partial}{\partial u^{+ i}}~,  \nn \\
&& \delta \zeta^M = \lambda^M(\zeta, u)~, \quad \delta u^{+}_i =
\lambda^{++}(\zeta, u)u^{-}_i~, \nn \\
&& \delta H^{++\,M} = {\cal D}^{++}\lambda^M -
\delta^{M}_{\mu+}\theta^{\mu+}\lambda^{++}~, \quad \delta H^{++++}
= {\cal D}^{++}\lambda^{++}~, \mu \equiv (\alpha, \dot\alpha)~, \nn \\
&&\delta {\cal D}^{++} = -\lambda^{++} D^0~, \quad D^0 \equiv
u^{+i}\frac{\partial}{\partial u^{+i}} - u^{-i}\frac{\partial}{\partial u^{-i}}
+ \theta^{\mu+} \frac{\partial}{\partial \theta^{\mu+}}~.
\eea
The vielbein
coefficients $H^{++M}, H^{++++}$ are unconstrained analytic superfields
involving an infinite number of the component fields which come from the
harmonic expansions. Most of these fields, like in $V^{++}$, can be gauged away by the
analytic parameters $\lambda^M, \lambda^{++}$, leaving in the WZ gauge just the
$(24+24)$ component fields of ${\cal N}=2$ Weyl multiplet. The invariant actions of various versions of
${\cal N}=2$ Einstein SG are given by a sum of the action of ${\cal N}=2$ vector
compensating superfield $H^{++ 5}(\zeta, u), \delta H^{++ 5} = {\cal
D}^{++}\lambda^5(\zeta, u)\,$, and that of matter compensator superfields, both
in the background of ${\cal N}=2$ conformal SG. The superfield $H^{++ 5}(\zeta, u)$
and extra gauge parameter $\lambda^5(\zeta,u)$ have, respectively, the
geometric meaning of the vielbein coefficient associated with an extra
coordinate $x^5$ (central charge coordinate) and the shift along this
coordinate \cite{SGEin}. It is assumed that nothing depends on $x^5$ . The most general
off-shell version of ${\cal N}=2$ Einstein SG is obtained by choosing the superfield
$q^{+a}(\zeta, u)$ as the conformal compensator. It involves an infinite number of
auxiliary fields and yields all the previously known off-shell versions with
finite sets of auxiliary fields via the appropriate superfield duality
transformations. Only this version allows for the most general SG-matter
coupling. The latter gives rise to a generic quaternion-K\"ahler (QK) sigma model in
the bosonic sector, in accordance with the theorem of Bagger and Witten
\cite{BW}. The general superfield Lagrangian of hypermultiplets in the background of ${\cal N}=2$
Weyl multiplet is a generalization of \p{33} to the SG case \cite{SGconf}, and it is the fundamental prepotential of
the quaternion-K\"ahler target geometry \cite{QK}. It can be used for the explicit computation of the QK metrics, e.g., through the appropriate
quotient construction in HSS \cite{BGIO,IVal}.

More references related to the basics of HSS can be found in the monograph \cite{book}.\\

\noindent{\bf 4.5 ${\cal N}=3$ harmonic superspace.}
The HSS method can be generalized to ${\cal N} > 2$. It was used to construct, for
the first time, an unconstrained off-shell formulation of ${\cal N}=3$ SYM
theory (that is equivalent to ${\cal N}=4$ SYM on shell) in the harmonic ${\cal N}=3$ superspace with
the purely harmonic part $SU(3)/[U(1)\times U(1)]$, $SU(3)$ being the
automorphism group of ${\cal N}=3, 4D$ supersymmetry \cite{4}. The corresponding action is written
in the analytic ${\cal N}=3$ superspace and has a nice form of the superfield
Chern-Simons term. This peculiarity supports the general statement  that the structure and geometry
of one or another gauge theory in superspace are radically different from those in the ordinary space-time.

Let us dwell on this formulation in some details. The ${\cal N}=3$ SYM constraints in the standard  ${\cal N}=3, 4D$
superspace read
\bea
&&\{{\cal D}^i_\alpha, {\cal D}^j_\beta\} = \varepsilon_{\alpha\beta} \bar{W}^{ij}\,, \quad
\{\bar{\cal D}_{\dot\alpha i}, \bar{\cal D}_{\dot\beta j}\} = \varepsilon_{\dot\alpha\dot\beta} \bar{W}_{ij}\,, \nn \\
&&\{{\cal D}^i_{\dot\alpha}, \bar{\cal D}_{\dot\beta j}\} = -2i\delta^i_j\,{\cal D}_{\alpha\dot\beta}\,, \lb{N32}
\eea
where $i, j = 1,2,3$ are indices of the fundamental representations of $SU(3)$ and $\bar{W}^{ij} = -\bar{W}^{ji}$ (together with its conjugate)
is the only independent covariant superfield strength  of the theory. Unlike the ${\cal N}=2$ SYM constraints,
eqs. \p{N32} put the theory on shell.

The basic steps in \cite{4} were the definition of the ${\cal N}=3$ harmonic superspace with the harmonic part
$SU(3)/[U(1)\times U(1)]$ parametrized by the mutually conjugated sets of harmonic variables possessing two independent harmonic $U(1)$ charges,
\bea
\Big(u^{(1,0)}_i,\, u^{(0,-1)}_i, \,u^{(-1,1)}_i\Big), \quad \Big(u^{i(-1,0)},\, u^{i(0,1)}, \,u^{i(1,-1)}\Big),
\quad u^{i(a,b)}u_i^{(c,d)} = \delta^{ac}\delta^{bd}\,,
\eea
and then the interpretation of the constraints \p{32} as the integrability conditions for the existence of an analytic subspace in such HSS:
\bea
\{{\cal D}_\alpha^{(1,0)}, {\cal D}_\beta^{(1,0)}\} = \{{\cal D}_\alpha^{(1,0)}, \bar{\cal D}_{\dot\beta}^{(0,1)}\}
= \{\bar{\cal D}_{\dot\alpha}^{(0,1)},
\bar{\cal D}_{\dot\beta}^{(0,1)}\} = 0\,,\lb{N3analit}
\eea
where ${\cal D}_\alpha^{(1,0)} = u^{(1,0)}_i{\cal D}^i_\alpha\,, \;\bar{\cal D}_{\dot\beta}^{(0,1)} = u^{i(0,1)}\bar{\cal D}_{\dot\alpha i}\,$.
The conditions
\p{N3analit} amount to the existence of a subclass of general ${\cal N}=3$ harmonic superfields,
the analytic superfields $\Phi^{(q_1, q_2)}(\zeta, u)$
living on the invariant analytic subspace with 8 independent Grassmann coordinates (as compared with 12 such coordinates
in the general ${\cal N}=3$ superspace),
\bea
\{\zeta, u\} = \{x^{\alpha\dot\alpha}_{\rm an},\, \theta^{(1,-1)}_\alpha,\,  \theta^{(0,1)}_\alpha,\,
\bar\theta^{(1,0)}_{\dot\alpha}, \,\bar\theta^{(-1,1)}_{\dot\alpha}, u \}.
\eea
The corresponding ${\cal N}=3$ Grassmann analyticity conditions are
\bea
{\cal D}_\alpha^{(1,0)}\Phi^{(q_1, q_2)} = \bar{\cal D}_{\dot\beta}^{(0,1)}\Phi^{(q_1, q_2)} = 0\,,
\eea
and they are solved as $\Phi^{(q_1, q_2)} = \Phi^{(q_1, q_2)}(\zeta, u)$ in the basis and frame in which the covariant
spinor derivatives ${\cal D}_\alpha^{(1,0)}$
and $\bar{\cal D}_{\dot\beta}^{(0,1)}$ simultaneously become ``short''. On the other hand, the triple of the harmonic derivatives
$\Big(D^{(2,-1)}, \,D^{(-1, 2)}, \,D^{(1,1)}\Big),$ which commute with ${\cal D}_\alpha^{(1,0)}, \bar{\cal D}_{\dot\beta}^{(0,1)}$
and so preserve the  ${\cal N}=3$
analyticity, acquire the analytic harmonic connections which are analogs of the ${\cal N}=2$ analytic gauge connection $V^{++}$:
\bea
\left(D^{(2,-1)}, D^{(-1, 2)}, D^{(1,1)}\right) \; \Rightarrow \; \left({\cal D}^{(2,-1)}, {\cal D}^{(-1, 2)}, {\cal D}^{(1,1)}\right), \;
{\cal D}^{(a, b)} = D^{(a, b)} + i V^{(ab)}(\zeta, u).
\eea
These harmonic derivatives satisfy, in both the original and the analytic bases, the commutation relations
\bea
[{\cal D}^{(2,-1)}, {\cal D}^{(-1, 2)}] = {\cal D}^{(1,1)}\,, \quad [{\cal D}^{(1,1)}, {\cal D}^{(2,-1)}] =
[{\cal D}^{(1,1)}, {\cal D}^{(-1,2)}] = 0\,.\lb{HarmHarm}
\eea

As was already mentioned, the constraints \p{N32} amount to the ${\cal N}=3$ SYM equations of motion and the same is
true for the equivalent form \p{N3analit}
of the same constraints. In the original basis the harmonic derivatives are short and their commutation relations with ${\cal D}_\alpha^{(1,0)}$
and $\bar{\cal D}_{\dot\beta}^{(0,1)}$,
\bea
&& [D^{(2, -1)}, {\cal D}^{(1,0)}_\alpha] = [D^{(-1, 2)}, {\cal D}^{(1,0)}_\alpha] = [D^{(1, 1)}, {\cal D}^{(1,0)}_\alpha] = 0\,, \nn \\
&& [D^{(2, -1)}, \bar{\cal D}^{(0,1)}_{\dot\alpha}] = [D^{(-1, 2)}, \bar{\cal D}^{(0,1)}_{\dot\alpha}] =
[D^{(1, 1)}, \bar{\cal D}^{(0,1)}_{\dot\alpha}] = 0\,, \lb{HarmGrass}
\eea
are satisfied for ${\cal D}_\alpha^{(1,0)}$ and $\bar{\cal D}_{\dot\beta}^{(0,1)}$ linearly depending on $SU(3)$ harmonics. Moreover, it can be shown
that \p{HarmGrass} are also the {\it necessary} conditions for ${\cal D}_\alpha^{(1,0)}$ and $\bar{\cal D}_{\dot\beta}^{(0,1)}$
to be linear in $SU(3)$ harmonics.
Thus the constraints \p{N32} are actually equivalent to the set of conditions \p{N3analit}, \p{HarmGrass} and \p{HarmHarm}
(with ${\cal D}^{(a,b)} = D^{(a,b)}\,$).

On the other hand, after solving \p{N3analit} by passing to the short ${\cal D}_\alpha^{(1,0)}\,, \bar{\cal D}_{\dot\beta}^{(0,1)}\,$,
and making the appropriate
similarity transformation of the remaining constraints, the relations \p{HarmGrass}
become the analyticity conditions for the three harmonic gauge connections $V^{(2,-1)}, V^{(-1, 2)}, V^{(1,1)}$ appearing
in the transformed harmonic derivatives. The whole dynamics proves to be concentrated in the purely harmonic constraints \p{HarmHarm} which are just the equations
of motion of the ${\cal N}=3$ SYM theory
in the analytic basis and frame. The final (and crucial) observation of ref. \cite{4} was that these equations can be reproduced
by varying the following Chern-Simons-type
off-shell analytic superfield action
\bea
S^{({\cal N}=3)}_{\rm SYM} &=& \int du d\zeta^{(-2, -2)} {\rm Tr} \Big\{ V^{(2,-1)}(D^{(-1, 2)}V^{(1,1)} -D^{(1, 1)}V^{(-1,2)}) \nn \\
&& -\, V^{(-1,2)}(D^{(2, -1)}V^{(1,1)} -D^{(1, 1)}V^{(2,-1)}) \nn \\
&& + \, V^{(1,1)}(D^{(2, -1)}V^{(-1,2)} -D^{(-1, 2)}V^{(2,-1)})\nn \\
&& - \, (V^{(1,1)})^2 + 2i V^{(1,1)}[V^{(2,-1)}, V^{(-1,2)}]\Big\}, \lb{N3CS}
\eea
where $du d\zeta^{(-2, -2)}$ is the appropriate integration measure over the analytic ${\cal N}=3$ superspace.
Like in the ordinary $3D$ non-abelian Chern-Simons action, varying \p{N3CS} with respect to the unconstrained analytic gauge potentials
yields the vanishing of three harmonic curvatures, which is equivalent to the relations \p{HarmHarm}. The off-shell invariance of the action \p{N3CS} under the
${\cal N}=3$ superconformal group $SU(2,2|3)$ has been shown in \cite{N3conF}.

The presence of just three harmonic gauge connections with three equations for them is only one reason for the existence
of an off-shell action for ${\cal N}=3$ SYM theory.
Two other reasons are the zero dimension of the integration measure of the ${\cal N}=3$ analytic superspace and the charge assignment
$(-2,-2)$ of this measure, which
precisely matches the zero dimension and the charge assignment $(2,2)$ of the analytic Lagrangian. This threefold coincidence
looks as a kind of ``miracle''.
Unfortunately, it fails to hold in the maximally extended ${\cal N}=4$ SYM theory. Though various harmonic superspace reformulations
of this theory were proposed
(see, e.g., \cite{5} where the ${\cal N}=4$ HSS with the harmonic part $SU(4)/[U(1)\times SU(2)\times SU(2)]$ was considered),
no any reasonable off-shell actions
were constructed in their framework so far. They merely serve to provide some new geometric interpretations
of the on-shell constraints of this theory. \\

Soon after its invention, the harmonic superspace approach was worldwide recognized as an adequate framework for
exploring theories with extended supersymmetry in diverse dimensions. Some of its further developments and uses are briefly outlined below. \\

\noindent{\bf 4.6 Quantum harmonic superspace.} The quantization of ${\cal N}=2$ theories in the harmonic formalism  was fulfilled  in \cite{Greenf}.
The actual applications of these quantum techniques started with the paper \cite{QuHSS} (see also the review \cite{32}) where
there was computed, for the first time,
the quantum one-loop effective action of the Coulomb phase of ${\cal N}=2$ SYM theory interacting with the massless and
massive matter hypermultiplets. The complete
agreement with the Seiberg-Witten duality hypothesis \cite{SeWit}  was found. The preservation of the manifest off-shell ${\cal N}=2$ supersymmetry
at all stages of computation was confirmed to be the basic advantage of the harmonic superspace quantum formalism. While in \cite{QuHSS} the effective action was
constructed in the sector of gauge fields, in the paper \cite{IKeZ} the analogous HSS-based one-loop computation was made in the hypermultiplet sector.
It was shown there that some non-trivial induced hyper-K\"ahler metrics (e.g., the Taub-NUT one) surprisingly come out as a quantum effect.

In \cite{BuI,BuI2}, we studied the issue of finding the leading term of the low-energy quantum effective action of ${\cal N}=4$ SYM theory
in the Coulomb phase in the ${\cal N}=2$ HSS formulation. In this formulation, the ${\cal N}=4$ SYM action is represented as a sum of the
${\cal N}=2$ SYM action and the action of the hypermultiplet in the adjoint representation minimally coupled to the ${\cal N}=2$
gauge potential $V^{++}$:
\begin{equation}
S^{({\cal N}=4)}_{\rm SYM}= S^{({\cal N}=2)}_{\rm SYM}
-\frac{1}{2} \mbox{Tr} \int du d \zeta^{(-4)} q^{+a}(D^{++}
+iV^{++}) q^+_a. \lb{N4actN2}
\end{equation}
Here $S^{({\cal N}=2)}_{\rm SYM}$ was defined in \p{Zupn} and $a=1,2$ is an index of the so called Pauli-G\"ursey group $SU(2)_{PG}$
which commutes with ${\cal N}=2$ supersymmetry. This combined action is invariant under the extra hidden ${\cal N}=2$ supersymmetry
\begin{equation}
\delta V^{++}=(\varepsilon^{\alpha a} \theta^+_a
+\bar{\varepsilon}^a_{\dot\alpha}\bar{\theta}^{+\dot\alpha})q^+_a,
\quad \delta q^+_a=-\frac{1}{2} (D^+)^4 \bigg[(\varepsilon^\alpha_a
\theta^-_\alpha +\bar{\varepsilon}_{\dot\alpha
a}\bar{\theta}^{-\dot\alpha})V^{--}\bigg]\label{offsec}
\end{equation}
(with $(D^+)^4 = \frac{1}{16}\,D^{+\alpha}D^+_\alpha\, \bar D^+_{\dot\alpha}\bar D^{+\dot\alpha}$),
which builds up the manifest ${\cal N}=2$ supersymmetry to ${\cal N}=4$ \footnote{Though \p{offsec}
is the symmetry of the off-shell action \p{N4actN2},
its correct closure with itself and with the manifest ${\cal N}=2$ supersymmetry is achieved only on shell.}.
The non-analytic gauge potential $V^{--}$ is related to $V^{++}$
by the harmonic flatness condition \p{harmflat}. In \cite{BuI}, based purely on the transformations \p{offsec},
we computed the leading term in the one-loop ${\cal N}=4$ SYM
effective action in the Coulomb phase (with the $SU(2)$ gauge group broken to $U(1)$) as
\be
\Gamma(V,q)=\frac{1}{(4\pi)^2}\int d^{12}z  \bigg\{\ln\frac{{\cal W}}{\Lambda}\ln\frac{\bar{\cal W}}{\bar \Lambda} +  \mbox{Li}_2(X) +\ln(1-X)
-\frac{1}{X}\ln(1-X)\bigg\}, \lb{EffWq}
\ee
where $\Lambda$ is an arbitrary scale, $X=\frac{-2q^{ai}q_{ai}}{{\cal W}\bar{\cal W}}$ and $\mbox{Li}_2(X)$ is the Euler dilogarithm.
In this formula, ${\cal W}$ is the chiral $U(1)$ superfield
strength and $q^{ia}$ is related to the on-shell $U(1)$ component of $q^{+a}$ as $q^{+a} = q^{ia}u^+_i$. Before \cite{BuI},
only the ${\cal W}$ part of \p{EffWq} was exactly known. The result \p{EffWq} was reproduced from
the quantum ${\cal N}=2$ supergraph techniques in \cite{BuI2}.

The quantum calculations in ${\cal N}=4$ SYM theory with making use of the harmonic ${\cal N}=2$ quantum supergraph techniques
are widely performed by other groups,
in particular, for checking the AdS$/$CFT correspondence (see, e.g., \cite{31} and references therein).\\

\noindent{\bf 4.7 Harmonic approach to the target geometries.} The fact that the general harmonic analytic Lagrangians of the hypermultiplets
in the rigid and local ${\cal N}=2$ supersymmetries can be identified with the prepotentials of the target space hyper-K\"ahler (HK)
and quaternion-K\"ahler (QK)
geometries was proved in \cite{HK,QK}. The general HK and QK constraints can be solved quite analogously to those of ${\cal N}=2$ SYM
or conformal SG theories, by
passing to $SU(2)$ harmonic extensions of the HK and QK manifolds and revealing there the appropriate analytic subspaces the dimension of which
is twice as less compared to that of the manifold one started with. The HK and QK constraints prove to admit a general solution
in terms of unconstrained
prepotentials defined on these analytic subspaces, and they are just the hypermultiplet Lagrangians mentioned above. The hypermultiplets $q^{+ n}$
are none other than the coordinates of these analytic subspaces. This deep affinity between the target and Grassmann harmonic
analyticities in the ${\cal N}=2, 4D$
(or ${\cal N}=4, 2D$) sigma models in the HSS approach looks very suggestive and surely deserves the further study and understanding.
The examples of such
an interplay  between the two types of the analyticity were also found for more complicated target geometries. For instance,
in a recent paper \cite{DeldIvNonl} the so called
HKT (``hyper-K\"ahler with torsion'') geometries (both ``weak'' and ``strong'' HKT) \cite{HKT}  were shown to select, as their  natural prepotentials,
the objects appearing
in the description of the most general $1D$ multiplets $({\bf 4, 4, 0})$ by the ${\cal N}=4, 1D$ analytic harmonic superfields \cite{ILe}
constrained by the
further harmonic conditions. One of the prepotentials  arises as the superfield Lagrangian of the $({\bf 4, 4, 0})$ analytic superfields,
while the other one
as a function defining the most general harmonic constraint for these superfields.\\

\noindent{\bf 4.8 Harmonic superspaces in diverse dimensions.} In \cite{IS,I1} the
bi-harmonic superspace with two independent sets of $SU(2)$ harmonics was
introduced and shown to provide an adequate off-shell description of ${\cal N}=(4,4),
2D$ sigma models with torsion. The analogous bi-harmonic ${\cal N}=4, 1D$ superspace \cite{INiedBi} secures the natural uniform description
of the models of ${\cal N}=4$ supersymmetric mechanics with the simultaneous presence of the ``mutually mirror'' worldline ${\cal N}=4$
multiplets. The harmonic superspace approach to extended supersymmetries in three dimensions was the subject
of the important papers \cite{Zu3,ZU2,Zrec}.
As a recent contribution in this direction, the ${\cal N}=3, 3D$ harmonic superspace formulation of the conformally
invariant ABJM (Aharony-Bergman-Jafferis-Maldacena)
theories was given in \cite{ABJM,ABJM2}. The harmonic superspace description of ${\cal N}=(1,0), 6D$ gauge theories and hypermultiplets
was worked out in \cite{Zu6,HStWe,ISZ,ISmi} and recently has received a further prospective development in \cite{BISmi}.
Various applications of
the harmonic superspace method in one-dimensional mechanics models and integrable systems are presented in \cite{ILe} and \cite{DI0} - \cite{DI1},
as well as in \cite{ILeFed}, \cite{IKoSmi}. In particular, ${\cal N}=4$, $1D$ HSS was used in \cite{DI0} to
construct ${\cal N}=4$ super KdV hierarchy. It was argued in \cite{DI} that the ${\cal N}=4, 1D$ harmonic superspace provides a unified
description of all known off-shell multiplets of ${\cal N}=4$ supersymmetric mechanics. The corresponding
${\cal N}=4, 1D$ superfields are related to each other via gauging the appropriate isometries of the superfield actions by
non-propagating ``topological'' ${\cal N}=4$ gauge multiplets.

Some other important
applications of the HSS approach involve classifying ``short''
and ``long'' representations of various superconformal groups in diverse
dimensions in the context of the AdS/CFT correspondence \cite{33}, study of the
domain-wall solutions in the hypermultiplet models \cite{AIN}, description of
self-dual supergravities \cite{DO}, construction of ${\cal N}=3$ supersymmetric Born-Infeld theory \cite{N3}, etc.
The Euclidean version of ${\cal N}=2$ HSS was
used in \cite{ILZ,FILSZ,ILZ2} to construct string theory-motivated
non-anticommutative (nilpotent) deformations of ${\cal N}=(1,1)$ hypermultiplet and
gauge theories.

By now, the HSS method has  proved its power as the adequate approach to off-shell theories
with extended supersymmetries. Without doubts, in the future it will remain the efficient and useful tool of dealing with
such theories.
\setcounter{equation}{0}

\section{Other related domains}
Here we briefly outline some other results obtained in the Sector 3 after the invention of supersymmetry.\\

\noindent{\bf 5.1 $2D$ integrable systems with extended supersymmetry.} In \cite{N2L} there was constructed,
for the first time, ${\cal N}=2$ supersymmetric
extension of the renowned $2D$ Liouville equation and the superfield Lax pair for it was found, as well as the general solution in a superfield
form. There was established, independently of \cite{GHL}, the existence of the twisted chiral representation of ${\cal N}=2, 2D$ supersymmetry
besides the standard chiral one. The method used in this construction was based on a nonlinear realization of infinite-dimensional ${\cal N}=2$
superconformal group in two dimensions, augmented with the inverse Higgs effect. Later on, the ${\cal N}=2$ Liouville equation appeared
in many contexts, including the ${\cal N}=2, 2D$ quantum supergravity closely related to string theory.

This research activity was continued in \cite{IK1}, where the same nonlinear realization methods were applied to the ``small'' ${\cal N}=4, 2D$
superconformal group to construct the new integrable superfield system,  ${\cal N}=4$ supersymmetric Liouville equation. Both the Lax representation
and general ${\cal N}=4$ superfield solution of this system were found. The ${\cal N}=4$ super Liouville equation is written as an equation for
the superfield describing the ${\cal N}=4, 2D$ ``twisted chiral'' multiplet and encompasses in its bosonic sector, along with the Liouville equation,
also the equations of Wess-Zumino-Novikov-Witten (WZNW) sigma model for the group $SU(2)$. So the system constructed simultaneously
yielded the first example of ${\cal N}=4$
supersymmetric extension of the WZNW sigma models playing the fundamental role in string theory and $2D$ conformal field theory\footnote{It was also the historically
first example of system with the target ``strong'' HKT geometry \cite{HKT}.}.

As a next development in the same direction, in \cite{IKL1} new ${\cal N}=4$ superextensions of WZNW sigma models were found, in particular
those exhibiting invariance under the ``large'' ${\cal N}=4, 2D$ superconformal groups. The relevant superfield and component
actions were presented and
it was shown that these systems admit deformations which preserve the original ${\cal N}=4$ superconformal symmetry and generate Liouville
potential terms in the actions.
In this way, new simultaneous superextensions of the Liouville equation and WZNW sigma models come out. The ${\cal N}=4, 2D$ WZNW sigma models
at the quantum level were studied in \cite{IKL2}.

A different sort of ${\cal N}=4$ supersymmetric integrable system was discovered in \cite{DI0}. It is an ${\cal N}=4$ superextension
of the KdV hierarchy.
Before this paper, only ${\cal N}=1$ and ${\cal N}=2$ supersymmetric KdV systems were known. The second Hamiltonian structure
of the new system was shown to be the
small ${\cal N}=4$ superconformal algebra with a central charge. The basic object is  the doubly charged harmonic analytic superfield subjected
to a simple harmonic constraint. Later on, there appeared a lot of papers devoted to further integrable extensions of this system and
their applications
in many mathematical and physical problems.\\

\noindent{\bf 5.2 Supersymmetric and superconformal mechanics.} The supersymmetric quantum mechanics \cite{Witt} is the simplest ($1D$)
supersymmetric theory. The first work on the extended superconformal
mechanics in the nonlinear realization superfield approach
was the paper \cite{IKL3}. There, the ${\cal N}=4$ superconformal mechanics associated
with the multiplet $({\bf 2, 4, 2})$\footnote{Such a notation for the off-shell multiplets
of $1D$ supersymmetry was suggested by A. Pashnev and F. Toppan in \cite{PaTo}. For the ${\cal N}=4, 1D$ case $({\bf n, 4, 4 -n})$
denotes a multiplet with 4 fermions, ${\bf n}$ physical bosons and ${\bf (4 -n)}$
auxiliary fields.} was reproduced and the new model with
the multiplet $({\bf 1, 4, 3})$ was found. Also, a new kind of the on-shell ${\cal N}$ extended superconformal mechanics
with the internal symmetry group $U({\cal N})$
and ${\cal N}$ fermionic fields in the fundamental representation of this group was constructed. The methods used in \cite{IKL3}
are based on the inverse Higgs phenomenon
which in this case has not only kinematic consequences, giving rise to the elimination of certain Goldstone superfields
in terms of few basic ones, but also
yields the dynamics, implying the equations of motion for the basic superfields. Results and methods developed in this
pioneer paper are actively applied and
developed in the studies related to the superconformal quantum mechanics, including the corresponding version of the AdS/CFT correspondence.
The closely related paper
is \cite{IKP}, where the phenomenon of partial breaking of ${\cal N}=4, 1D$ supersymmetry was studied for the first time,
on the example of the multiplet $({\bf 1, 4, 3})$.

As other benchmarks on the way of developing this line of research it is worth to distinguish the papers \cite{ILe} and \cite{ILeFed}.

In \cite{ILe}, the harmonic superspace method was adapted to $1D$ supersymmetric models,
{\it i.e.} the models of supersymmetric quantum mechanics, and then applied for constructing the superfield actions of diverse
${\cal N}=4, 1D$  multiplets, including
the sigma-model type actions, superpotentials and the superfield Wess-Zumino (or Chern-Simons) terms. The realization
of the most general ${\cal N}=4, 1D$
superconformal group $D(2,1;\alpha)$ in the $1D$ harmonic superspace was found and a wide class of new models of
supersymmetric (and superconformal) ${\cal N}=4$ mechanics
was constructed. This paper triggered many subsequent papers of different authors on the related subjects.

In \cite{ILeFed}, new superconformal
extensions of integrable $1D$ Calogero-type models were constructed by gauging the $U(n)$ isometries of matrix superfield models
(with the use of methods of refs. \cite{DI}). The cases of
${\cal N}=1, 2,$ and  ${\cal N} = 4$ superconformal systems were considered. The ${\cal N}=4$ extension of the so called ``$U(2)$ spin''
Calogero system was deduced. The paper \cite{ILeFed} was
first to introduce the spin (or ``isospin'') superfield variables, with the WZ type action of the first order in the time derivative
for their bosonic physical components. In the subsequent studies, these variables proved to be a useful tool of constructing various new models
of ${\cal N}=4$ and ${\cal N}=8$
supersymmetric mechanics, including the models in which the ${\cal N}=4, 1D$ multiplets couple to the external
non-abelian gauge fields \cite{IKoSmi}.

The further developments along these lines with the participation of the Dubna group, together with the relevant references,
can be retrieved from the reviews \cite{ISmiA} and \cite{FILrev}.  As a recent new direction of research, it is worth to mention
the deformed ${\cal N}=4$ mechanics associated with the supergroup $SU(2|1)$ \cite{Deformed}. The relevant models involve the intrinsic
mass parameter and go over to the standard ${\cal N}=4$ mechanics models, when this parameter goes to zero.

A different approach was represented by the papers \cite{IMT1,IMT2,IMT3} and \cite{FuzI,FuzIBych},
in which the target-space supersymmetrization of the quantum-mechanical Landau problem on a plane and two-sphere was treated, as well as the
closely related issue of ``fuzzy'' supermanifolds. In some cases, the worldline supersymmetry  arises as a hidden symmetry of such models.
These studies look rather interesting and perspective, since, e.g., they are expected to give rise to a deeper understanding
of quantum Hall effect and its possible superextensions. The relationships of these models to superparticles and
superbranes are also worthy to learn in more depth.
\\

\noindent{\bf 5.3 Superparticles, branes, Born-Infeld, Chern-Simons,  and higher spins.} In the end of nineties, there was growth of interest in the
superfield description of superbranes as systems realizing the concept of Partial Breaking of
Global Supersymmetry (PBGS) pioneered by Bagger and Wess \cite{BW1} and Hughes and Polchinsky \cite{HP}.
In this approach, the physical worldvolume superbrane
degrees of freedom are represented by Goldstone superfields, on which the
worldvolume supersymmetry acts by linear transformations. The
rest of the full target supersymmetry is spontaneously broken and is realized nonlinearly.
In components, the transverse coordinates of the superbrane (if they exist) are
described by a gauge-fixed Nambu-Goto action.
In the cases when the Goldstone supermultiplets are vector ones, the Goldstone superfield actions
simultaneously provide supersymmetrization of the appropriate Born-Infeld-type
actions. The relevant references can be found, e.g., in \cite{I,II}.

Among the most important results obtained in this domain with the decisive participation of the Dubna group
it is worth to mention the perturbative-theory construction of
the ${\cal N}=2$ superfield Born-Infeld action with the spontaneously broken ${\cal N}=4$ supersymmetry \cite{BIK,N4N2},
as well as the interpretation of a hypermultiplet as a Goldstone multiplet supporting a partial breaking of ${\cal N}=1,\, 10D$
supersymmetry  \cite{10susy}. The peculiarities of the partial breaking ${\cal N}=2\,\rightarrow \, {\cal N}=1$ in $4D$ within
the ${\cal N}=2$ superfield formalism (including the interplay between the electric and magnetic Fayet-Ilopoulos terms)
were discussed in \cite{IvZup}. The ${\cal N}=3$ supersymmetric extension of the Born-Infeld theory was constructed in \cite{N3}.

In \cite{AdStran}, the so called ``AdS/CFT equivalence transformation'' was proposed. It relates  the standard realization of the spontaneously
broken $4D$  conformal group $SO(2,4)$ on the dilatonic field (``conformal basis'') with its realization as the isometry group
of the gauge-fixed $AdS_5$  brane (``AdS basis''). The $1D$ version of this transformation allowed us to show \cite{NiedIK}
that the standard one-dimensional conformal mechanics  is in fact equivalent to the so called ``relativistic conformal mechanics''
of ref. \cite{RelCM} (alias $AdS_2$ particle). This  correspondence can be extended to superconformal mechanics models in the
Hamiltonian formalism \cite{AdSHam} and is now widely applied in many domains (see, e.g., \cite{Crem}).

Supersymmetric extensions of the Chern-Simons terms in three-dimensions (as well as of their generalization, the so called $BF$
Lagrangians) were constructed and studied in \cite{Zu3} - \cite{ABJM2} and \cite{ZuPak} - \cite{ZupBF}. In particular, in \cite{ZuPak}
and \cite{IChSi} the manifestly supersymmetric superfield form of the ${\cal N}=2$ Chern-Simons action was given for the first time.

In \cite{GoyIv}, it was shown that the $AdS_3 \times S^3$ and $AdS_5 \times S^5$ superstring theories in the Pohlmeyer-reduced form \cite{Pohl} reveal hidden
${\cal N}=(4,4)$ and ${\cal N}=(8,8)$ worldsheet supersymmetries. The explicit form of the supersymmetry transformations was found,
for both the off-shell action and the superstring equations.

A new superfield approach to the higher-spin multiplets based on nonlinear
realizations of the generalized $4D$ superconformal group $OSp(1|8)$ has been developed in
\cite{ILu}. It was argued that the higher-spin generalization of ${\cal N}=1$ supergravity
should be based, {\it a la} Ogievetsky and Sokatchev,  on the preservation of the $OSp(1|8)$ analog of chirality.
There were also given a few proposals of how to reproduce the higher spin equations by quantizing
various kinds of superparticles \cite{Master,FILu,Maxw}. In particular, it was shown in \cite{Maxw} that a new kind of such equations
can be obtained by quantizing a particle in the tensorial space associated with the so called Maxwell extension of the Poincar\'e group.
The BRST approach to Lagrangian formulation of higher-spin fields was successfully elaborated
by A. Pashnev with co-authors
(see \cite{BuPash} and references therein).\\

\noindent{\bf 5.4 Last but not least: Auxiliary tensor fields for duality invariant theories.} Nowadays, the duality invariant systems
attract a lot of attention (see, e.g., \cite{KuTh} and references therein). The simplest example of
duality in $4D$ is the covariance of the free Maxwell equation and Bianchi identity,
\be
\partial^mF_{mn} = 0\,, \quad \partial^m\tilde{F}_{mn} =0\,, \quad   \tilde{F}_{mn} :=\frac12 \varepsilon_{mnpq}F^{pq}\,, \lb{FreeMax}
\ee
under the $O(2)$ duality rotation
\be
\delta F_{mn} = \omega \tilde{F}_{mn}\,, \quad \delta \tilde{F}_{mn} = -\omega {F}_{mn}\,, \lb{O2dual}
\ee
where $\omega$ is a real transformation parameter. Another example of the duality-invariant system is supplied by the renowned
nonlinear Born-Infeld theory. The duality invariant systems involving, besides gauge fields, also the coset scalar fields described by non-linear sigma models naturally
appear in various extended supergravities and are important ingredients of string/brane theory.

Even in the simplest $O(2)$ duality case it was not so easy to single out  the most general set of duality invariant nonlinear
generalizations of the Maxwell theory. There were  developed a few approaches based on solving some nonlinear equations. In \cite{IZuD,IZuRev}
a new purely algebraic approach to this problem was proposed. Namely, it was shown that the most general duality-invariant nonlinear extension
of Maxwell theory is described by the Lagrangian
\bea
&&{\cal L}(V,F)={\cal L}_2(V,F)+E(\nu,\bar{\nu}),\lb{GenDual} \\
&&{\cal L}_2(V,F)=\frac12(\varphi+\bar\varphi)+ \nu+ \bar{\nu} - 2\,(V\cdot F+\bar{V}\cdot\bar{F}),
\eea
where the auxiliary unconstrained fields $V_{\alpha\beta}$ and $\bar{V}_{\dot\alpha\dot\beta}$ were introduced, with $\nu=V^2,\; \bar\nu=\bar{V}^2\,$,
$\varphi = F^{\alpha\beta}F_{\alpha\beta}, \bar\varphi = \bar{F}^{\dot\alpha\dot\beta}\bar{F}_{\dot\alpha\dot\beta}$ and $F^{\alpha\beta},
\bar{F}^{\dot\alpha\dot\beta}$ representing the Maxwell field strength in the spinorial notation.
In \p{GenDual},  ${\cal L}_2(V,F)$ is the bilinear part only through which the Maxwell field strength enters
the action and $E(\nu, \bar\nu)$ is the nonlinear interaction involving only auxiliary fields. The duality group acts on $V_{\alpha\beta}$ as
$$
\delta V_{\alpha\beta}=-i\omega V_{\alpha\beta}, \qquad \delta\nu=-2i\omega\nu\,,
$$
and it was proved that the requirement  of duality invariance of the full set of equations of motion following from \p{GenDual} amounts to
$O(2)$ invariance of the function $E(\nu,\bar\nu)$,
\be
E(\nu,\bar\nu) = {\cal E}(a)\,, \quad a = \nu\bar\nu\,.
\ee
Eliminating the auxiliary fields from \p{GenDual} with such $E(\nu,\bar\nu)$ by their algebraic equations of motion, we obtain a nonlinear
version of Maxwell action, such that the relevant equations of motion necessarily respect duality invariance. Thus the variety of all possible duality invariant extensions
of the Maxwell theory is parametrized by the single function ${\cal E}(a)$ which can be chosen at will.

Later on, this formalism was generalized to the cases of $U(N)$ duality \cite{IZ2} and $Sp(2,\mathbb{R})$ duality \cite{ILZup1}. The ${\cal N}=1, 2$
superfield extensions were built in \cite{Kuz}-\cite{IZ4}.

Note that the tensorial auxiliary field representation was guessed from the construction
of ${\cal N}=3$ superfield Born-Infeld theory in \cite{N3}. These auxiliary fields naturally appear as the necessary components of the off-shell
${\cal N}=3$ SYM multiplet in the HSS approach. Keeping this in mind, it seems probable that the tensor auxiliary fields formulation
of the duality invariant systems could also enter as an element into the hypothetical harmonic superfield formulations of various extended supergravities. \\

Finally, it is worth mentioning that, besides the topics listed above, in Sector ``Supersymmetry'' for the last decade the investigations
on a few different important subjects were also accomplished. These include the twistor approach to strings and particles (see, e.g., \cite{Fedor,Fedor22}),
the studies related to the AGT (Alday-Gaiotto-Tachikawa) conjecture  (see, e.g., \cite{Mar}) and, more recently,
the explicit construction of instantons and monopoles (see, e.g., \cite{Yasha}). In view of lacking of space, I will not dwell on these
issues.

\section{Conclusions}
In this paper, I reviewed the mainstream scientific activity of the Sector of Markov - Ogievetsky - Ivanov -\,... for more than fifty
years. In retrospect, the most influential pioneering results and methods which have successfully passed the examination by time are, in my opinion,
the following: {\bf I.} Notoph; {\bf II.} ``Ogievetsky Theorem'' and the view of the gravitation theory as a theory of spontaneous breaking, with the graviton as
a Goldstone field; {\bf III.} The inverse Higgs phenomenon; {\bf IV.} The complex superfield geometry of ${\cal N}=1$ supergravity; {\bf V.}
The general relationship between linear and nonlinear
realizations of supersymmetry; {\bf VI.} Grassmann analyticity and harmonic superspace.

As for the future directions of research, I think that in the nearest years they will be mainly concerned with exploring the geometry and quantum structure of
supersymmetric gauge theories and supergravity in diverse dimensions in the superfield approach,
as well as studying various aspects of supersymmetric and superconformal mechanics models in their intertwining relationships with the higher-dimensional
field theories and string theory.

\vspace{5mm}\noindent
{\Large\bf Acknowledgements}\nopagebreak
\vspace{0.4cm}

\noindent
The author is grateful to the Directorate of BLTP for the suggestion to prepare this review and S. Fedoruk for useful remarks. The work
was partially supported by RFBR grants, projects No 15-02-06670 À and No 16-52-12012, and a grant
of the Heisenberg-Landau program.

\end{document}